\documentclass[aps,prc,twocolumn,superscriptaddress]{revtex4-2}
\usepackage{amssymb}
\usepackage{amsmath}
\usepackage{color}
\usepackage{graphicx}
\usepackage[hidelinks]{hyperref}
\def\be{\begin{equation}}
\def\ee{\end{equation}}
\def\bea{\begin{eqnarray}}
\def\eea{\end{eqnarray}}
\usepackage{ulem}
\usepackage{color} % to paint the words to be indexed

\usepackage{bbm}
\begin{document} 

%Title of paper
\title{Octupole correlations in superdeformed bands of  $^{56}$Ni} 

\author{Xiao Lu} 
\affiliation{CAS Key Laboratory of Theoretical Physics, Institute of Theoretical Physics, Chinese Academy of Sciences, Beijing 100190, China} 
\author{Shivani Jain} 
\affiliation{CAS Key Laboratory of Theoretical Physics, Institute of Theoretical Physics, Chinese Academy of Sciences, Beijing 100190, China} 
\author{Hiroyuki Sagawa} 
\email[]{hiroyuki.sagawa@gmail.com}
\affiliation{CAS Key Laboratory of Theoretical Physics, Institute of Theoretical Physics, Chinese Academy of Sciences, Beijing 100190, China}
\affiliation{RIKEN Nishina Center for Accelerator-Based Science, Wako 351-0198, Japan}
\affiliation{Center for Mathematics and Physics University of Aizu, Aizu Wakamatsu, Fukushima 965-0001, Japan} 
\author{Shan-Gui Zhou} 
\email[]{sgzhou@itp.ac.cn}

%\homepage[]{Your web page}
%\thanks{} 
\affiliation{CAS Key Laboratory of Theoretical Physics, Institute of Theoretical Physics, Chinese Academy of Sciences, Beijing 100190, China} 
\affiliation{School of Physical Sciences, University of Chinese Academy of Sciences, Beijing 100049, China} 
\affiliation{School of Nuclear Science and Technology, University of Chinese Academy of Sciences, Beijing 100049, China}  

\date{\today}

\begin{abstract} 
The projected multi-dimensionally-constrained relativistic Hartree-Bogoliubov model was employed to calculate the potential energy surface of the high-spin states in $^{56}\text{Ni}$. It is pointed out for the first time that possible octupole deformations exist for the positive and negative parity superdeformed bands in $^{56}\text{Ni}$, with deformations $\beta_{30}\sim0.14$ and $\beta_{30}\sim0.24$, respectively,  along with a large prolate deformation of $\beta_{20}\sim 0.42$. 
These octupole deformations are induced by the coupling between $2p_{3/2}$ and $1g_{9/2}$ orbits at the deformation $\beta_{20}\sim 0.4$. 
The calculated excitation energies of the two rotational bands are consistent with the observed superdeformed bands of $^{56}\text{Ni}$. In addition, two rotational bands are predicted, consisting of one superdeformed band with negative parity and one hyperdeformed bands with positive parity.
\end{abstract}

% insert suggested keywords - APS authors don't need to do this
%\keywords{}

%\maketitle must follow title, authors, abstract, and keywords
\maketitle
\section{Introduction} 
The ground states of medium-heavy and heavy nuclei, especially in the rare-earth and actinide regions $150<A<190$ and $A>220$, exhibit spatially reflection-symmetric quadrupole deformation, often associated with hexadecapole deformation. However, in specific regions of the nuclear chart, the reflection symmetry is spontaneously broken, and pear-like octupole deformation appears in the ground state due to the Jahn-Teller effect \cite{Butler96,Butler20PRL}.   
Possible octupole deformation is induced by the coupling  between the normal parity single-particle state specified by $(l,j)$ and the unique parity single-particle state with $(l',j')$, which satisfy the conditions $\Delta l=|l-l'|=3$ and $\Delta j=|j-j'|=3$. This requirement is met for proton or neutron numbers being close to 34 ($1g_{9/2} \leftrightarrow 2p_{3/2}$ coupling), 56 ($1h_{11/2}\leftrightarrow 2d_{5/2}$ coupling), 88 ($1i_{13/2}\leftrightarrow 2f_{7/2}$ coupling), and 134 ($1j_{15/2}\leftrightarrow 2g_{9/2}$ coupling) \cite{Butler96}.  The numbers 34, 56, 88, and 134 are called ``octupole magic numbers".

Experimental evidence for octupole deformation has been found in the mass region around $A\sim 220$, $150$ and $80$ \cite{Butler2016,Butler20,han23,Butler_2024}. For instance, in the light actinide nuclei with $(N,Z)\simeq(134,88)$, such as $^{220}$Rn \cite{RN949}, $^{224}$Ra \cite{224Ra20}, and  $^{228}$Th \cite{RN950}, octupole deformation has been observed. In the mass region $(N,Z)\simeq (88,56)$, both $^{144}$Ba \cite{144Ba16} and $^{146}$Ba \cite{146Ba17} have been found to exhibit octupole deformation in the intrinsic frame.  
Thus, the regions around $A\sim 220$ and $150$ are well-established as ``islands of octupole deformation", while the area around $A \sim 80$ is relatively new, with discoveries primarily in isotopes with $34\leq Z\leq 38$ \cite{han23,2023IndicationsOO,Butler_2024}, except for $^{71}$Ge \cite{ge712022} and $^{67}$Ga \cite{ga672024}. However, octupole correlation effects in nuclei with $Z<34$, especially around $A\sim 60$, have not been extensively explored, making it an intriguing topic for further study.

Strong octupole correlations may occur in not only the ground state but also the excited states. In 1999, two superdeformed rotational bands were observed in $^{56}$Ni, one of which is a positive-parity rotational band (SD1), while the other (SD2) has an unspecified parity \cite{RN619}. A linking transition between these two bands has been identified \cite{RN619,RN616,RN620}. Both experimental and theoretical studies suggest that the SD2 band has negative parity \cite{RN619,RN616}.  Notably, the cranked Hartree-Fock and Hartree-Fock-Bogolyubov calculations describe the formation of the SD2 band as the excitation of a proton or neutron from the $[321]1/2$ orbit to the $[440]1/2$ orbit, corresponding to the excitation from the spherical orbits $2p_{3/2}$ to $1g_{9/2}$, which can be influenced by octupole correlations \cite{RN619}. The above information inevitably leads to the speculation that the two rotational bands in $^{56}$Ni could be a pair of positive and negative parity bands associated with octupole correlations. 
Although a large number of theoretical studies have provided detailed analysis of the collectivity in $^{56}$Ni \cite{RN633,honma02prc,RN629,RN630,RN632,RN652,RN654,RN644,RN636}, the possible octupole correlation has still been overlooked.  

Theoretically, many approaches have been developed to describe and predict octupole correlations and deformation in atomic nuclei, such as the macroscopic-microscopic model \cite{NPA1984,MOLLER2008758,Kowalprc17,RN942}, microscopic mean-field or many-body methods using various interactions \cite{Dobaczewskiprl05,Robledoprc11,Robledo_2012,LI2013866,luprc14,Zhou_2016,Wangk2022,Chenprc15,Agbemavaprc16,Nomuraprc18,Ganev19,Lizp19,cyc20,hext20,Nomura21,Nomuraprc22,Guojyprc22,RONG2023,Minhprc24,Nomuraprc24,wang24,zhaoymprc24,XU2024138893}, the cluster model \cite{Shneidmanprc03,Buck2008}, and other elaborated models \cite{Zamfirprc01,Lu_2018,WANG2019454,Minkov22,qibin22,wangyyprc22,Jolosprc12,Antonenkoprc23,Jolosprc2024,MS_2024,Minkov2024}. Among these theoretical models, multidimensionally constrained covariant density functional theory (MDCCDFT) is recognized as an effective framework within a self-consistent mean-field approach \cite{luprc14,luprc2012r,Wangk2022}.
This model allows for the simultaneous consideration of reflection-asymmetric and axially asymmetric deformations, requiring the spatial $V_4$ symmetry, which permits only even $\mu$ value for the multipole deformation $(\lambda,\mu)$  \cite{Wangk2022}. 
Depending on how to deal with the pairing correlation, two variants were developed: the multidimensionally constrained relativistic mean-field model \cite{luprc14,luprc2012r}, and the multidimensionally constrained relativistic Hartree–Bogoliubov model (MDCRHB) \cite{zhaoprc2017}. MDCCDFT has been extensively used to study normal and hypernuclei, focusing on properties related to various deformation effects on nuclear characteristics. These studies encompass fission barriers of actinide nuclei \cite{luprc14,luprc2012r}, the third minima in potential energy surfaces (PESs) of light actinides \cite{zhaoprc2015}, shapes and PESs of superheavy nuclei \cite{meng2019-m}, nonaxial octupole $Y_{32}$ correlations \cite{zhaoprc2012-m,zhaoprc2017}, axial octupole $Y_{30}$ correlations \cite{XU2022137287,RN418,WANG2019454,Chen021301,Wang044316,ge712022}, and the structure of hypernuclei \cite{Lu014328,Lu044307,RONG2020135533,Rong054321,RN956,Sun6153}.

However, one of the most salient characteristics of mean-field approaches is that the solutions often spontaneously break symmetries of the Hamiltonian, making it impossible to label the system with symmetry quantum numbers such as angular momentum, parity, etc. To cure these shortcomings, it is necessary to consider effects beyond mean-field models, such as angular momentum and the parity projections. 
These projections can be implemented using the Generator Coordinate Method (GCM), projected Random Phase Approximation, or any models that accommodate deformations and projection techniques. Recently, a projected MDCRHB (p-MDCRHB) model was developed by incorporating parity and angular momentum projections into the MDCRHB model to study the low-lying excited states associated with exotic nuclear shapes, including octupole deformation \cite{Wangk2022,RONG2023}. In this study, we employ both the MDCRHB and p-MDCRHB models to investigate octupole correlations in the superdeformed bands of $^{56}$Ni.

The paper is organized as follows, Section \ref{sec2} provides a general introduction of the adopted method. The results and discussion are presented in Section \ref{sec3}. The summary and future perspectives are given
in Section \ref{sec4}.

\section{Theoretical framework}\label{sec2}
The details of the MDCRHB and p-MDCRHB theory with point-coupling density functional can be found in Refs. \cite{luprc14,Zhou_2016,Wangk2022}. Here, we only present briefly the formalism for the convenience of the following discussions. In MDCRHB model, the RHB equation in coordinate space reads
\begin{equation}\label{RHB}
\begin{aligned}
\int \mathrm{d}^3 \boldsymbol{r}^{\prime}\left(\begin{array}{cc}
h-\lambda & \Delta \\
-\Delta^* & -h+\lambda
\end{array}\right)\binom{U_k}{V_k}  =E_k\binom{U_k}{V_k},
\end{aligned}
\end{equation}
where $\Delta$ and $\lambda$ are the pairing and chemical potential, respectively. $E_k$ is the quasi-particle energy and $\left(U_k(\mathbf{r}),V_k(\mathbf{r})\right)^T$ is the quasi-particle wave function. $h$ is the single particle Hamiltonian
\begin{equation}
h=\boldsymbol{\alpha} \cdot \boldsymbol{p}+\beta[M+S(\boldsymbol{r})]+V(\boldsymbol{r}),
\end{equation}
where $M$ is the nucleon mass, $S(\boldsymbol{r})$ and  $\boldsymbol{V}(\boldsymbol{r})$ are scalar and  vector potentials, respectively. We use the Bogoliubov transformation together with a separable pairing force of finite range \cite{TIAN200944,PRC09tian}
\begin{equation}
V=-G \delta\left(\boldsymbol{R}-\boldsymbol{R}^{\prime}\right) P(\boldsymbol{r}) P\left(\boldsymbol{r}^{\prime}\right) \frac{1-P_\sigma}{2}
\end{equation}
to treat the pairing effect. Here, $G$ is the pairing strength, $\boldsymbol{R}=\left(\boldsymbol{r}_1+\boldsymbol{r}_2\right) / 2$ and $\boldsymbol{r}=\boldsymbol{r}_1-\boldsymbol{r}_2$ are the center of mass and relative coordinates, respectively. The symbols with and without the prime denote the quantities before and after the interaction, respectively. $P_\sigma$ is the spin-exchange operator. $P(\boldsymbol{r})=\left(4 \pi a^2\right)^{-3 / 2} e^{-r^2 / 4 a^2}$ is the Gaussian function.  The values of pairing strength $G=728.0$ MeV$\cdot$fm$^{3}$ and the effective range of the pairing force $a=0.644$ fm are used in the  present  calculations. All these parameters are obtained by fitting to the pairing gap in the nuclear matter \cite{TIAN200944,PRC09tian-1}.

In this work, we use the PC-PK1 parameter set \cite{Zhaopw10} and solve the RHB equation by expanding the wave function in an axially-deformed harmonic oscillator (ADHO) basis. The resulting minima in PES represent the ground state or the shape isomers. To obtain the PESs, a modified linear constraint method is used, i.e., the Routhian with deformation constraints reads
\begin{equation}
E^{\prime}=E_{\mathrm{MF}}+\sum_{\lambda \mu} \frac{1}{2} C_{\lambda \mu} \beta_{\lambda \mu},
\end{equation}
where the variable $C_{\lambda \mu}$ varies during the iteration. $\beta_{\lambda \mu}$ is the nuclear deformation 
\begin{equation}
\beta_{\lambda \mu}=\frac{4 \pi}{3 A R^\lambda}\left\langle Q_{\lambda \mu}\right\rangle ,
\end{equation}
where $R=1.2A^{1/3}$ fm is the nuclear radius parameter, $A$ is the number of nucleons. The multipole moment operator is given by $Q_{\lambda\mu}=r^{\lambda}Y_{\lambda\mu}(\Omega)$, where $Y_{\lambda\mu}$ is the spherical harmonic. Under the imposed symmetry in
this model, all $\beta_{\lambda \mu}$ with even $\mu$ can be considered simultaneously.

To obtain the properties of excited states, the angular momentum projection and parity projection were applied in the p-MDCRHB model \cite{Wangk2022}. The projected wave function reads 
\begin{equation}
\left|\Psi_{\alpha, q}^{J M \pi}\right\rangle=\sum_K f_\alpha^{J K \pi} \hat{P}_{M K}^J \hat{P}^\pi|\Phi(q)\rangle,
\end{equation}
where $f_\alpha^{J K \pi}$ is the weight function, $q$ represents a collection of the deformation parameters. 
%$|\Phi(q)\rangle$ is the wave function obtained by MDCRHB calculations. 
 $\hat{P}_{M K}^J$ and $\hat{P}^\pi$ are angular momentum and parity projection operators, respectively. The $f_\alpha^{J K \pi}$ coefficient and the corresponding energy $E_\alpha^{I\pi}$ are determined via the Hill-Wheeler-Griffin equation \cite{yao09}
\begin{equation}\label{HWG}
\sum_{K^{\prime}}\left[\mathcal{H}_{K K^{\prime}}^{I \pi}-E_\alpha^{I \pi} \mathcal{N}_{K K^{\prime}}^{I \pi}\right] f_{K^{\prime} \alpha}^{I \pi}=0
\end{equation}
where $\mathcal{H}_{K K^{\prime}}^{I \pi}=\langle\Phi| H P^\pi P_{K K^{\prime}}^I|\Phi\rangle$ and $\mathcal{N}_{K K^{\prime}}^{I \pi}=\langle\Phi| P^\pi P_{K K^{\prime}}^I|\Phi\rangle$ represent the Hamiltonian kernel and the norm overlap kernel, respectively.

Although the mean-field intrinsic states are obtained with the constraint on the correct particle number on average, this does not guarantee the correct particle number in the angular momentum projected states \cite{yao09}. To cure this problem, one has to perform a particle number projection (PNP) calculation together with the angular momentum and parity projections. However, the projection calculations to restore both the rotational and parity symmetry breakings are already high computational burden, and further PNP calculations would exacerbate these computational circumstances. Thus, in this work, we choose to handle PNP  by a Lipkin-Nogami-type model adding two particle number constraint terms to the Hamiltonian kernel to approximately conserve the proton and neutron numbers 
as described in Ref. \cite{yao044311}.

\section{Results and discussions}\label{sec3}
\subsection{PES and deformed single-particle levels}
\begin{figure}
 \centering
 \includegraphics[width=8cm]{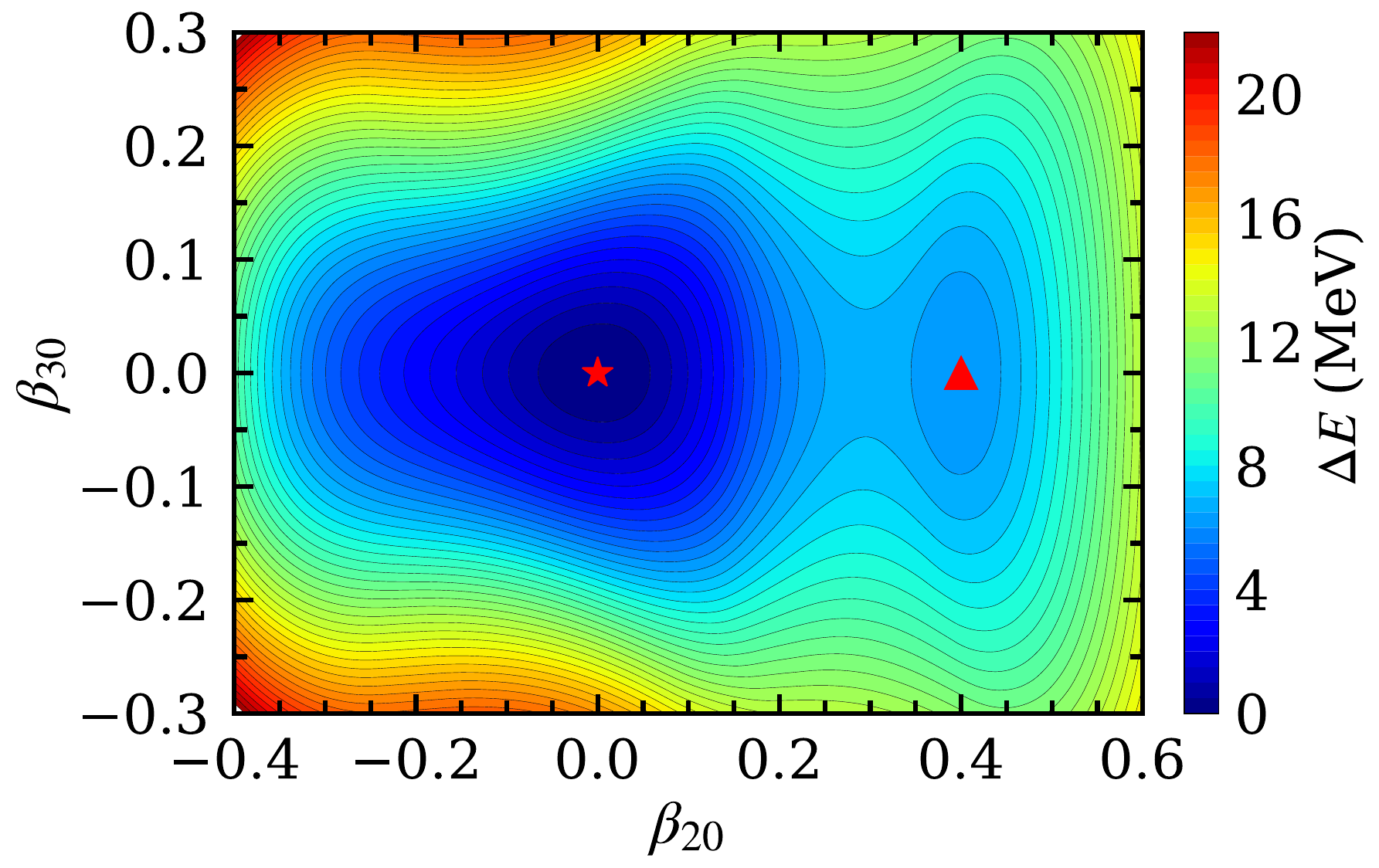}
 \caption{\label{MF-PES}(Color online) The potential energy surface for $^{56}$Ni in the $\beta_{20}$-$\beta_{30}$ plane from the constrained MDCRHB calculations with density functional PC-PK1. The ground state and the second minimum are labeled by star and triangle, respectively. The energy is normalized with respect to the binding energy of the ground state. Each contour line represents an energy separation of 0.5 MeV.
}
\end{figure}   
In MDCRHB and p-MDCRHB, both reflection and axial symmetries can be broken, and the shape degrees of freedom $\beta_{\lambda\mu}$, where $\mu$ is an even number, are self-consistently included, such as $\beta_{20}$, $\beta_{22}$, $\beta_{30}$, $\beta_{32}$, $\beta_{40}$, $\beta_{42}$, and $\beta_{44}$. In the following study, only the reflection symmetry is broken, while the axial symmetry is preserved for convenience, i.e., we take $\beta_{22}=\beta_{32}=\beta_{42}=\beta_{44}=0$.  For $^{56}$Ni, we truncate the expansion of the ADHO basis at $N_f=14$, and the potential and density are calculated on a spatial lattice, where the number of mesh points in the $\rho$ and $z$ directions are set to 12 and 24, respectively, ensuring the convergence of the energy.

We first calculated the two-dimensional PES of $^{56}$Ni using MDCRHB with density functional PC-PK1. As both $\beta_{20}$ and $\beta_{30}$ are constrained in the calculation of PES, it can be seen in Fig. \ref{MF-PES} that $^{56}$Ni has a spherical ground state and a prolate local minimum, represented by a red star and a red triangle, respectively. The PES is very soft with respect to the shape degree of freedom $\beta_{30}$, which is very similar to the PES of $^{78}$Br in the $A\approx80$ mass region where the octupole correlation has been observed  \cite{RN418}. This similarity suggests that $^{56}$Ni may also exhibit strong octupole correlations. 

%\sout{Therefore, we speculate that} 
The PES in Fig. \ref{MF-PES} suggests that the two experimentally  observed superdeformed bands have a large prolate deformation. This speculation is supported by the  dynamical quadrupole moment $Q_{20}$ extracted from the reduced transition probability $B(E2)$ values of the inband transitions of SD2 members, which corresponds to a quadrupole deformation of $\beta_{2}=0.37$ \cite{RN619,RN618}. This value is very close to our calculated prolate minimum $\beta_2=0.40$.   
\begin{figure}
 \centering
 \includegraphics[width=8cm]{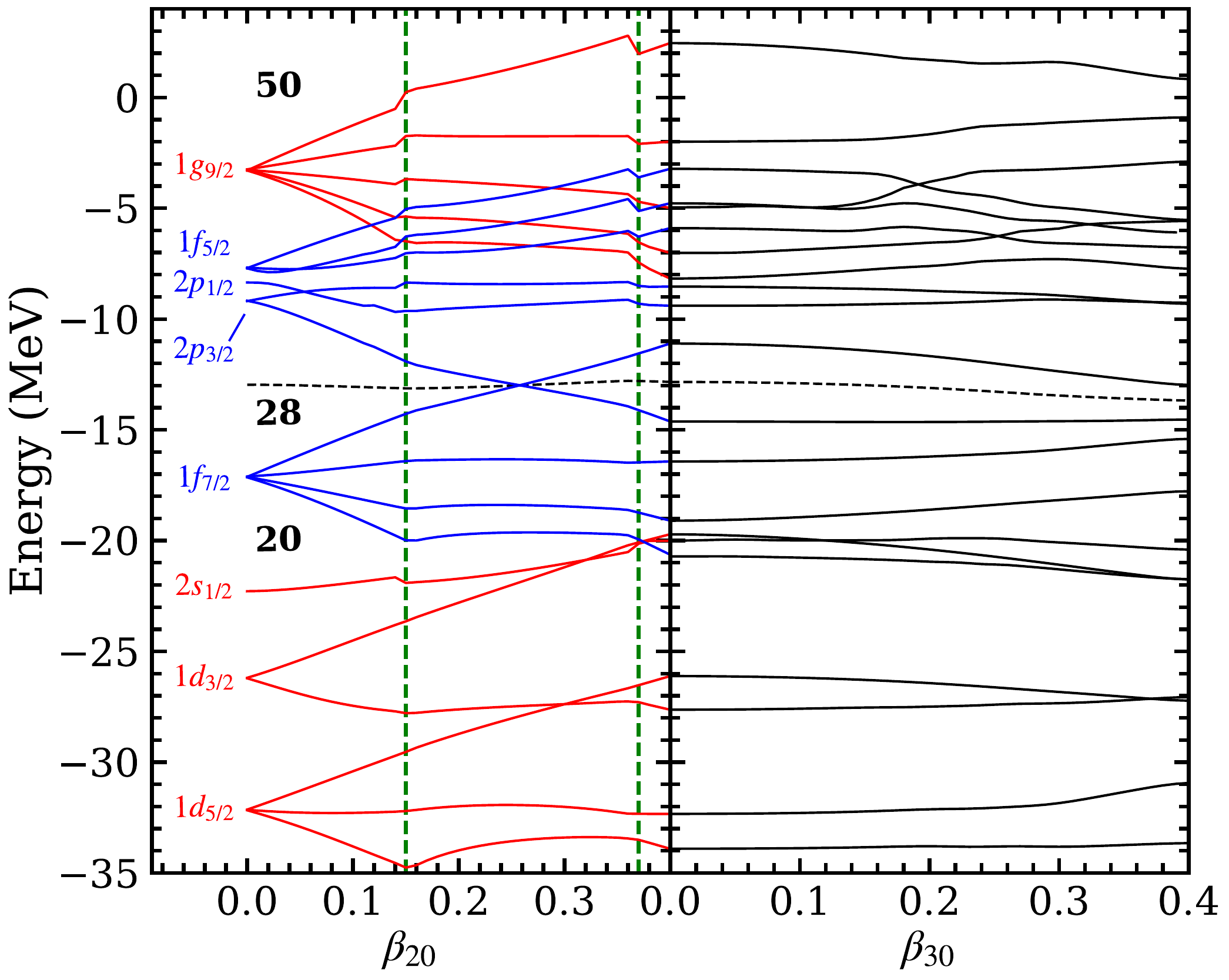}
 \caption{\label{neufer}(Color online) Single-neutron levels of $^{56}$Ni as functions of $\beta_{20}$ (left panel) and $\beta_{30}$ (right panel) deformations. The positive and negative parities in the left panel are denoted by red and blue, respectively. The principal, orbital, and angular momentum quantum numbers for spherical symmetry are presented. The black dash line denotes the Fermi level. In the left panel, the pairing energy is non-zero in the region between the green dashed lines and zero in all other areas.
}
\end{figure}

We also presented the single-neutron levels as functions of $\beta_{20}$ and $\beta_{30}$ deformations in Fig. \ref{neufer}. From this figure, it can be seen that as the quadrupole deformation increases,  
the $2p_{3/2}$ orbital gradually evolves below the Fermi surface. Notably, at $\beta_{20}= 0.4$, two nucleons occupy the $2p_{3/2}$ orbital. Moreover, as the $\beta_{20}$ changes, the $2p_{3/2}$ and $1g_{9/2}$ orbitals get closer to each other, leading to increased configuration mixing between them. At this point, the nucleons have a probability of being excited from the $2p_{3/2}$ orbital below the Fermi surface to the $1g_{9/2}$ orbital above the Fermi surface, thereby generating significant octupole correlation effects in the superdeformed states in $^{56}$Ni. %\sout{within the nucleus.}   
We should notice that the present MDCRHB  results indicate no  $\beta_{30}$ minimum, but  exhibit relative softness in the $\beta_{30}$ direction, which suggests possible octupole deformation in the excited states. The changes in single-partcle neutron orbitals are rather small as $\beta_{30}$ increases at the fixed $\beta_{30}$=0.4.

\begin{figure*}
 \centering
 \includegraphics[width=16cm]{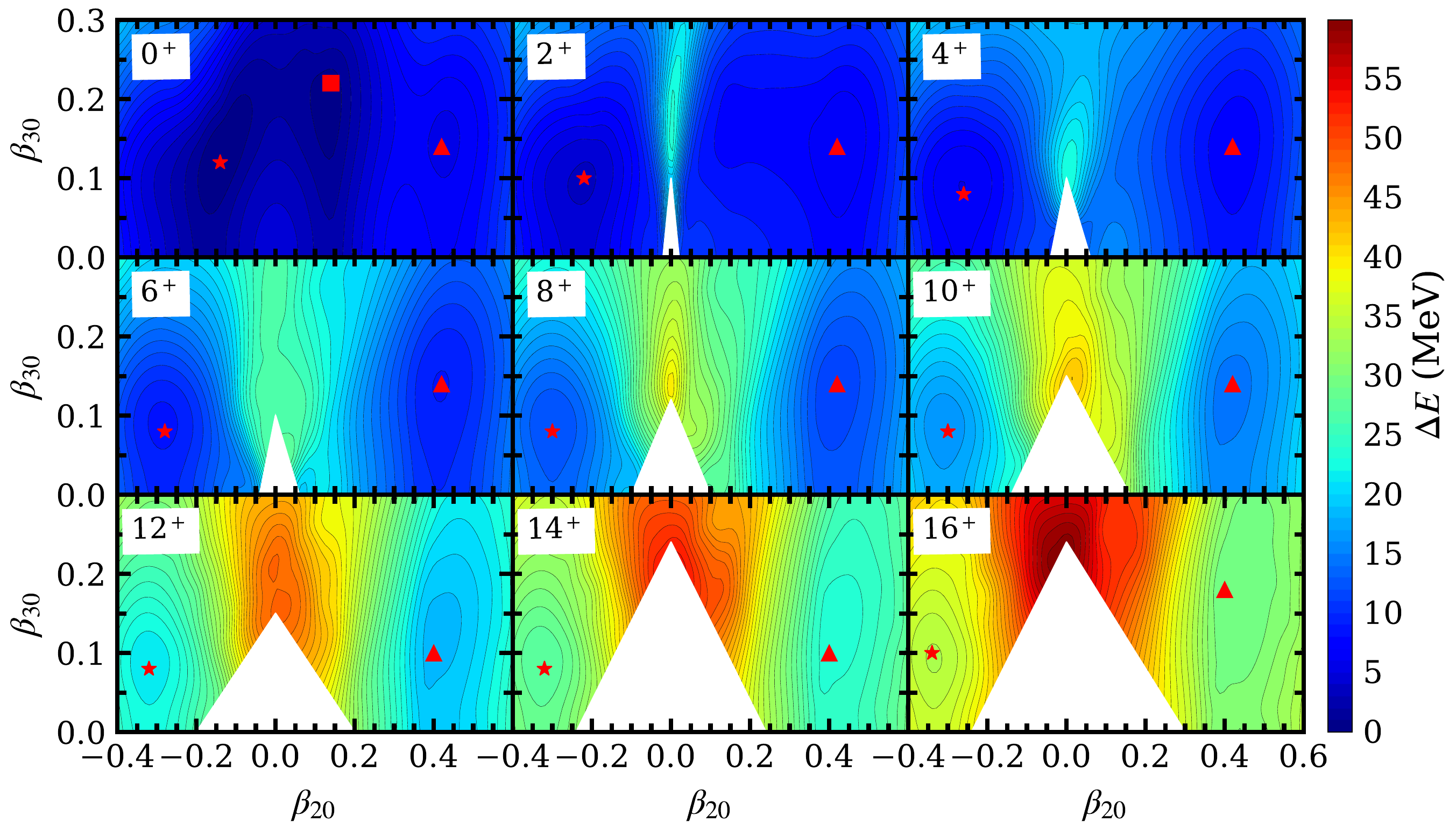}
 \caption{\label{pes-pos}(Color online) The potential energy surfaces of the projected $J^\pi=0^+,2^+,4^+,6^+,8^+$, $10^+,12^+,14^+$ and $16^+$ states in $^{56}$Ni calculated by using p-MDCRHB approach with density functional PC-PK1. All energies are referred  to the projected ground state ($0_1^+$). Yrast states and superdeformed states with positive parity are labeled by stars and triangles, respectively. The $0_2^+$ state is indicated by a square. Each contour line represents an energy separation of 1 MeV. %\blue{(HS: need to explain what are blank regions in figures.)}
The blank region near $\beta_{20}=0$ for $J>0$ is omitted  in the projected PES because of numerical instability problem. See the text for details.}
\end{figure*}  

\begin{figure}
 \centering
 \includegraphics[width=8cm]{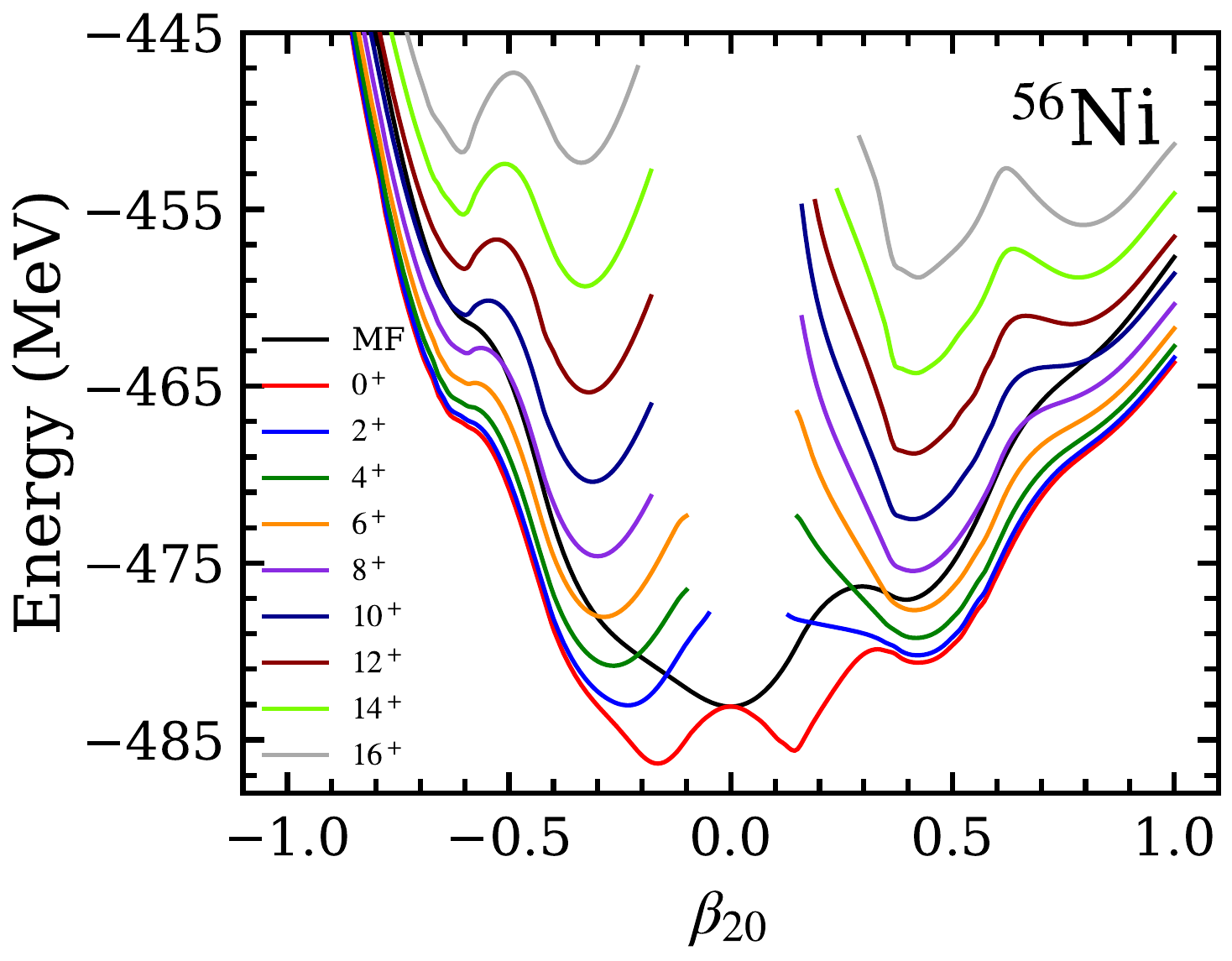}
 \caption{\label{energycurve-pos}(Color online) The potential energy curves of the projected positive parity states $0^+$-$16^+$ in $^{56}$Ni, as functions of the quadrupole deformation parameter $\beta_{20}$. The intrinsic deformed states are obtained by the constrained MDC-RHB calculations with density functional PC-PK1.  
}
\end{figure}
\subsection{Positive parity states  of  p-MDCRHB results}
Based on the calculation results of the PESs at the mean-field level, we further perform angular-momentum and parity projections using the p-MDCRHB method to obtain the two-dimensional projected PESs for positive parity states from $0^+$ to $16^+$. The results are presented in Fig. \ref{pes-pos}, where the energy of the projected PESs is defined as $\Delta E =E_{\text{proj}}-E(0^+)_{\text{min}}$. Here, $E_{\text{proj}}$ and $E(0^+)_{\text{min}}$ represent the projection energy and global energy minima in the $0^+$ PES, respectively. 
Notice that the values near the spherical point $\beta_{20}=0$ (shown by blank triangle region) are discarded in the projected PESs for $J\geq1$ because of numerical instability problem: for the states with $J\neq 0$,  the quantities of the Hamiltonian kernel $\mathcal{H}_{K K^{\prime}}^{I \pi}$ and/or the norm overlap kernel $\mathcal{N}_{K K^{\prime}}^{I \pi}$  around $\beta_{20}\approx 0$ in Eq. (\ref{HWG}) are so small that their ratio cannot be determined accurately.  
These configurations can be safely omitted from the projected energy curves of $J>0$ states since the intrinsic spherical shape may not play a role for high angular momentum states \cite{RN769}.

From Fig. \ref{pes-pos}, it is observed that for the $0^+$ state, the projected PES exhibits three minima.  
The ground state $0_1^+$ is no longer spherical but rather has a slight oblate deformation  $\beta_{20}=-0.14$ together with octupole deformation $\beta_{30}=0.12$. The second minimum $0_2^+$ near the spherical shape corresponds to a deformation $\beta_{20}=0.14$, $\beta_{30}=0.22$ with an excitation energy of $0.17$ MeV, while the third minimum $0_3^+$ corresponds to a deformation $\beta_{20}=0.42$, $\beta_{30}=0.14$ with an excitation energy of $5.69$ MeV. The energy of the third minimum is very close to the experimentally observed energy 5 MeV of the third $0^+$ state \cite{ni561974}. Other positive parity states also exhibit minima on their PESs that are close to the deformation value of the $0_3^+$ state as seen in Fig. \ref{pes-pos}. For clarity, we show in Fig. \ref{energycurve-pos} the potential energy curves (PECs) for the mean-field (MF) and the positive parity states from $0^+$ to $16^+$, obtained by constraining only the $\beta_{20}$ deformation, with axial symmetry and reflection asymmetry maintained in the calculations. It can be clearly seen from the deformation and the excitation energy of $0_3^+$ state 
that this state corresponds to the bandhead of the SD1 band. 
Furthermore, we have plotted the relative excitation energies corresponding to these minima, i.e., $E_x(J^\pi)=E(J^\pi)_{\text{min}}-E(0^+)_{\text{min}}$, in Fig. \ref{energycom} and compared them with the experimental values of the SD1 band. It can be seen that our calculated results correspond reasonably well with the experimental results, especially in the low spin states $J\leq 6$. It is worth noting that the experiments did not provide the $0^+$ state in the SD1 band; however, based on our calculations, the experimentally observed third $0^+$ with the excitation energy of 5 MeV is highly likely to belong to this band. 

Additionally, in the projected PESs from $2^+$ to $16^+$, there are minima in the oblate deformation region for $\beta_{20}$ values from $-0.22$ to $-0.34$ and octupole deformation around $\beta_{30}\sim 0.1$, which may correspond to the yrast band observed experimentally. The relative excitation energies corresponding to the minima are also shown in Fig. \ref{energycom}, where it can be seen that the theoretically calculated band structure is consistent with the experimental data, although the calculated moment of inertia is smaller than that of the experimental data. Currently, the wave functions considered are obtained from the projection of single mean-field wave function, which may be insufficient to accurately describe high-spin states. To better describe high spin states, further consideration of configuration mixing, such as multireference covariant density-functional theory \cite{RN769,yao044311,zhao041301,sun064319,zhou034305}, is desperately needed. 
 
Besides, as can be seen in Fig. \ref{energycurve-pos}, the PECs as a function of $\beta_{20}$ were expanded in the range of $-1.0\leq\beta_{20}\leq1.0$. In addition to the yrast and SD1 bands mentioned above, a hyperdeformed rotational band also appears in the deformation regions of the oblate shape with $\beta_{20} \sim -0.6$, which can be referred to as HD1 band. The corresponding excitation energies are also shown in Fig. \ref{energycom}. It is worth noting that minima are also observed in the high-spin states $12^+$, $14^+$, and $16^+$ for the prolate shape with $\beta_{20} \sim 0.8$. However, the band structure is not well defined, and further analysis will not be conducted in this study.

\subsection{Negative parity states  of  p-MDCRHB results}

\begin{figure*}[ht!]
 \centering
 \includegraphics[width=16cm]{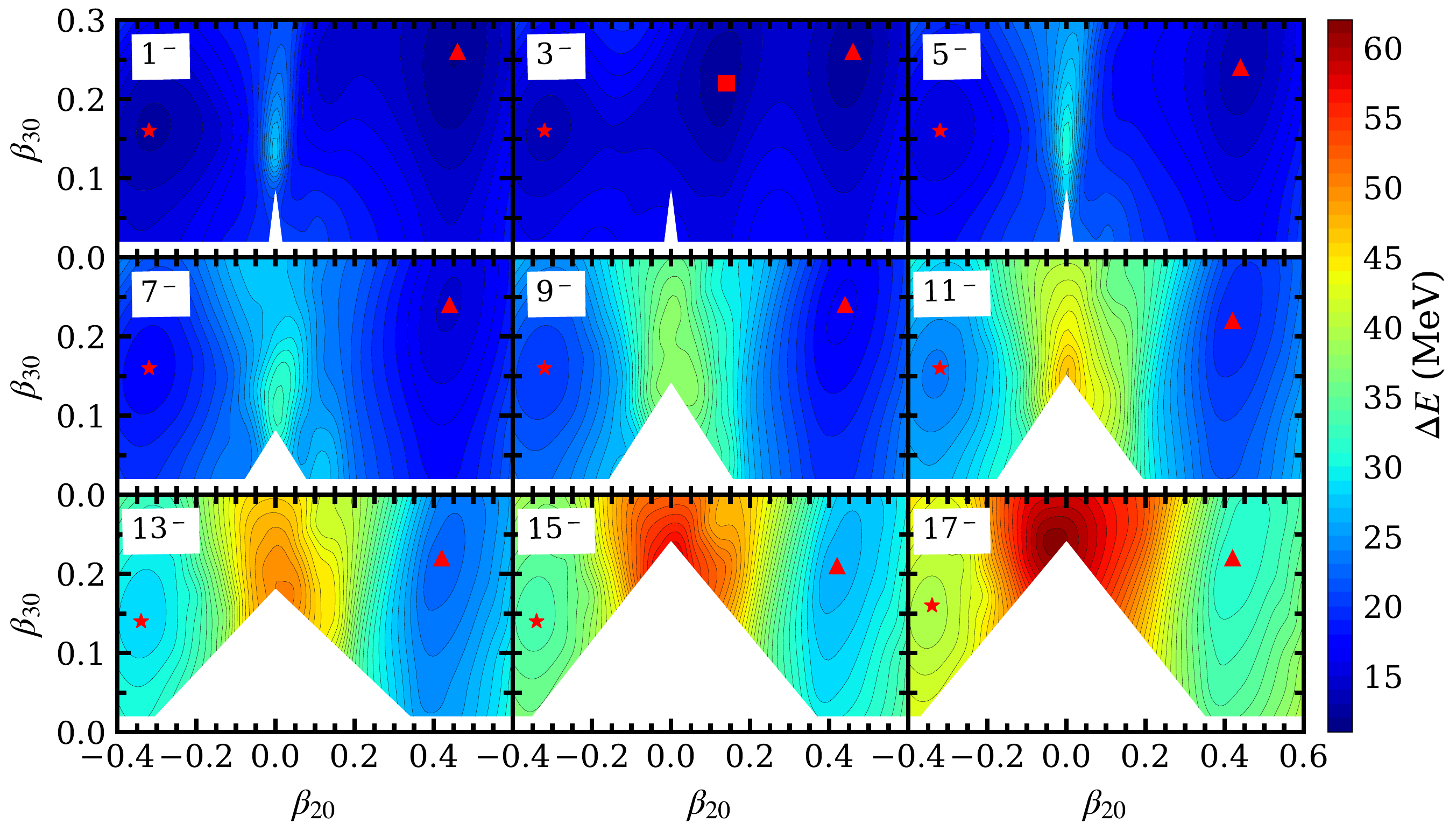}
 \caption{\label{pes-neg}(Color online) The potential energy surfaces of the projected $J^\pi=1^-,3^-,5^-,7^-,9^-$, $11^-,13^-,15^-$ and $17^-$ states in $^{56}$Ni calculated by using p-MDCRHB approach with density functional PC-PK1. All energies are referred  to the projected ground state ($0_1^+$). The superdeformed states with negative parity and $\beta_{20}<0$ are labeled with stars, whereas those with $\beta_{20}>0$ are labeled with triangles. The $3_1^-$ state is indicated by a square. Each contour line represents an energy separation of 1 MeV. The blank region near $\beta_{20}=0$ for $J>0$ is omitted  in the projected PES because of numerical instability problem. See the text for details.
}
\end{figure*} 

\begin{figure}
 \centering
 \includegraphics[width=8cm]{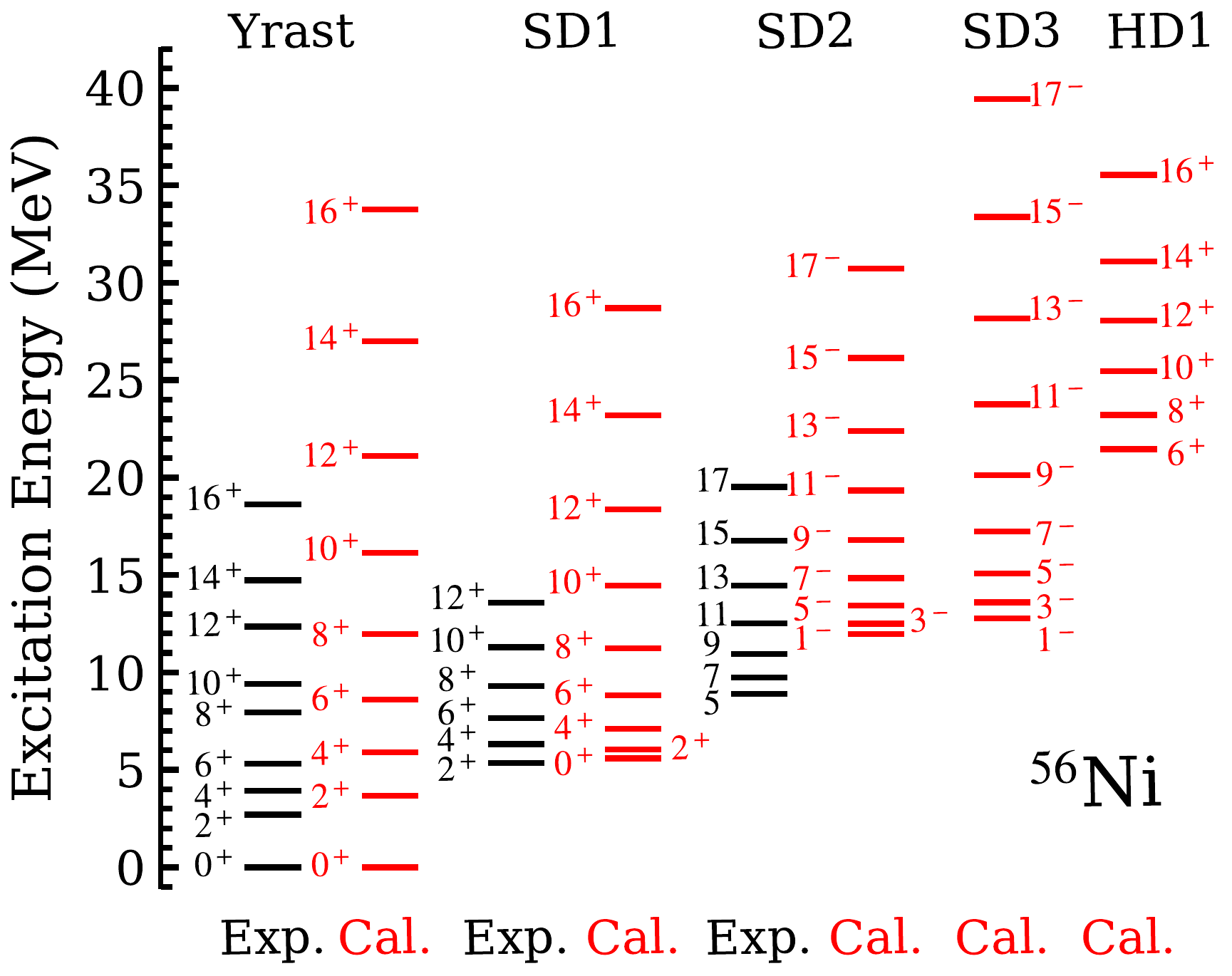}
 \caption{\label{energycom}(Color online) Calculated excitation energies in $^{56}$Ni obtained by using p-MDCRHB approach with density functional PC-PK1 compared to the experimental values. The experimental values are taken from Refs. \cite{RN619,RN620}.
}
\end{figure}

Next, one can employ the p-MDCRHB calculations to obtain the projected PESs for the negative parity states from $1^-$ to $17^-$, as shown in Fig. \ref{pes-neg}. Except for the $3^-$ state, the PESs for the other spins states have two minima. The minima in the prolate deformation region correspond to $\beta_{20}\sim 0.42$, $\beta_{30}\sim 0.24$, while the minima in the oblate deformation region are around $\beta_{20}\sim -0.34$, $\beta_{30}\sim 0.16$.  This could be related to the two regular rotational bands. According to the observed connecting transitions between the SD2 band and the SD1 band \cite{RN620}, we assign that the minima at the prolate deformation, which is similar to the deformation value of the SD1 band, correspond to the experimentally observed SD2 band. On the other hand, the minimum associated with oblate deformation may represent another unobserved superdeformed band with higher energies in $^{56}$Ni, referred to as the SD3 band. The excitation energies for the SD2 and SD3 bands are also shown in Fig. \ref{energycom}. In contrast to the comparative analysis of the theoretically and experimental SD1 band, the calculated SD2 band predicts a higher energy than the experimental data. This is likely due to the insufficiency of the current wave function in describing the negative parity superdeformed bands, similar to the issue with the positive parity superdeformed band. 
\begin{figure}
 \centering
 \includegraphics[width=8cm]{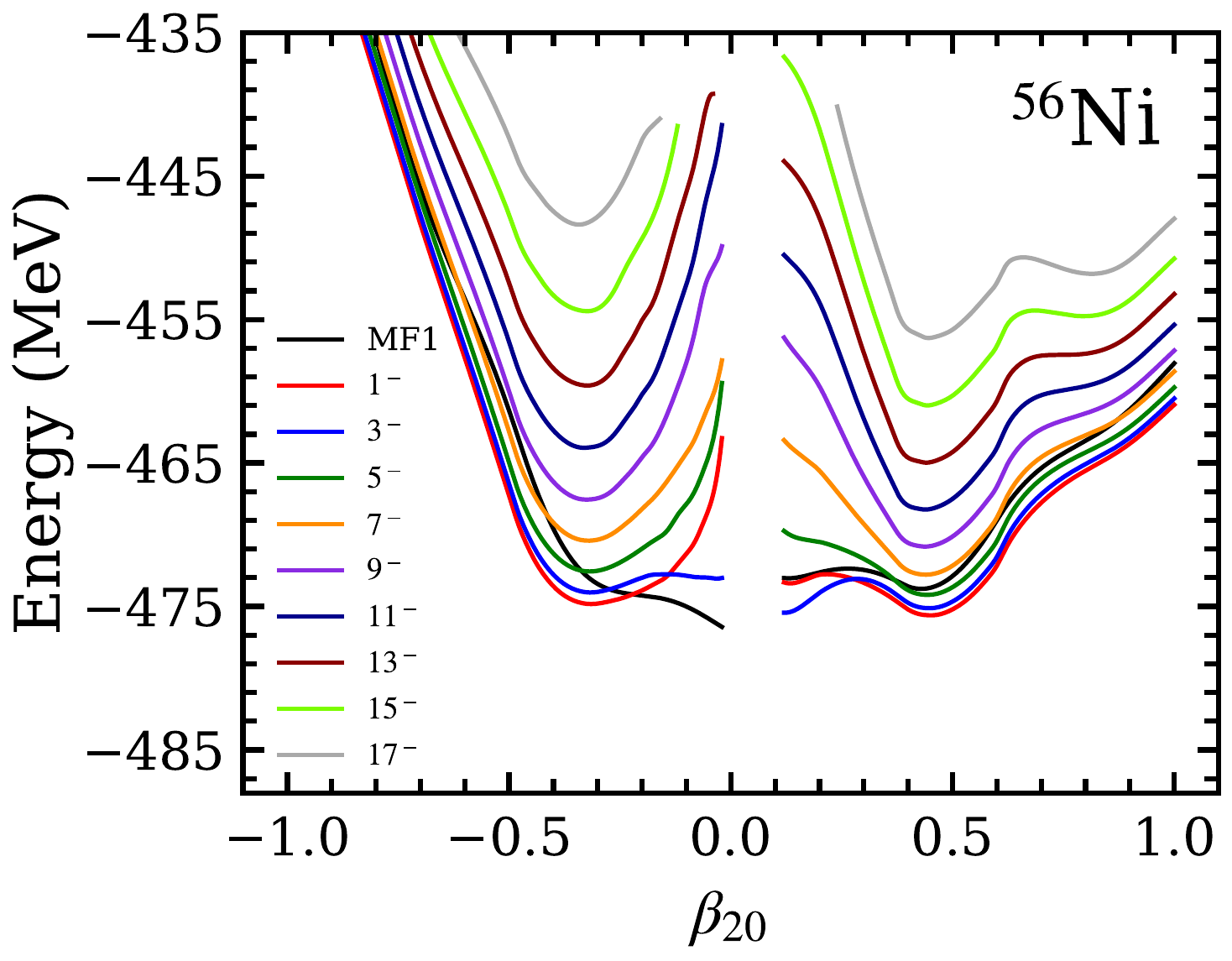}
 \caption{\label{energycurve-neg}(Color online) The potential energy curves of the projected negative parity states $1^-$-$17^-$ in $^{56}$Ni, as functions of the quadruple deformation parameter $\beta_{20}$. The intrinsic deformed states are obtained by constrained MDC-RHB calculations with density functional PC-PK1. In the figure, $\beta_{30}$ is constrained to $0.16$ for $\beta_{20} < 0$ and to $0.24$ for $\beta_{20} > 0$ in the calculation.
}
\end{figure}

Similar to Fig. \ref{energycurve-pos}, we have also expanded the deformation range of $\beta_{20}$ and performed the PECs calculation using p-MDCRHB for the negative-parity states from $1^-$ to $17^-$, as shown in Fig. \ref{energycurve-neg}. To ensure the reflection asymmetry of the system, we fixed $\beta_{30}=-0.16$ for the oblate deformation  and $\beta_{30}=0.24$ for the prolate deformation. In addition to the SD2 and SD3 bands, minima are also observed in the high-spin states $15^-$ and $17^-$.
%a candidate of  hyperdeformed band also appears near $\beta_{20} \sim 0.8$ in the calculated results. 

\section{Summary}\label{sec4}
Strong octupole correlations may occur not only in the ground state but also in the excited states. In 1999, the two superdeformed rotational bands SD1 and SD2 have been observed in $^{56}$Ni. This study primarily investigates the possible octupole correlations in these two rotational bands. 

We have first investigated the two-dimensional PES of $^{56}$Ni within the MDCRHB framework. The results show that there are two minima in $^{56}$Ni, with deformation values of $\beta_{20}=0.0$ and $\beta_{20}=0.4$. The PES is very soft with respect to the shape degree of freedom $\beta_{30}$, which is very similar to the case of $^{78}$Br in the $A\approx80$ mass region where the octupole correlation has been observed \cite{RN418}. The local minimum at $\beta_{20}=0.4$ is proposed as the deformation associated with the two superdeformed bands.
Furthermore, single-neutron level analysis shows that as $\beta_{20}$ changes, the $2p_{3/2}$ orbit goes down below the Fermi energy and has a large occupation probability. At the same time, $1g_{9/2}$ orbit goes down and get closer to $2p_{3/2}$ orbit. These features in the deformed single-particle diagram may enhance the configuration mixing and induce significant octupole correlation effects in the superdeformed bands of $^{56}$Ni.

The projected PESs for the positive parity states $0^+$ to $16^+$ and the negative parity states $1^-$ to $17^-$ in $^{56}$Ni were studied with  the p-MDCRHB model. The results suggest for the first time that the two superdeformed bands, SD1 and SD2, in $^{56}$Ni correspond to a pair of octupole-deformed bands; the positive-parity band SD1 has the deformations $\beta_{20}=0.42$ and $\beta_{30}=0.14$, while the negative-parity band SD2 has the deformations  $\beta_{20}=0.42$, and $\beta_{30}=0.24$. The experimentally observed third $0^+$ state at 5 MeV is suggested to belong to the SD1 band, as its energy is very close to the value calculated in our study, with $E(0^+_3)=5.69$ MeV having almost the same deformation as the SD1 band.

An oblate deformation from $-0.14$ to $-0.34$ associated with the  octupole deformation around $\beta_{30}\sim 0.1$ is suggested for the yrast  band by the  p-MDCRHB calculations. The calculated  energies of the low-spin members of yrast and SD1 bands are in reasonable agreement with the experiment data, while the calculated energies of the higher spin members of the yrast and SD1 bands, and also thoses of the SD2 band show higher energies  compared to the experimental data. This is likely due to the insufficiency of the adopted wave function in describing the higher spin states of positive parity bands and negative parity band. To better describe these states, we should take into account  the configuration mixing effects based on multi Slater determinants with different deformations in p-MDCRHB framework in the future.

In addition, a superdeformed rotational band SD3 with deformation at $\beta_{20} \sim -0.34$, $\beta_{30} \sim 0.16$ is predicted in the projeced PESs of the negative parity states, and its excitation energy is higher than that of the SD2 band. Further calculations of the one-dimensional PECs also predict a positive parity hyperdeformed band with deformation $\beta_{20}\sim -0.6$ and several minima in the high spin states with deformation $\beta_{20}\sim 0.8$.
Experiments are highly desired to confirm the prediction of the octupole rotational bands SD1 and SD2, the superdeformed band SD3, and the hyperdeformed bands in $^{56}$Ni.

\begin{acknowledgments} 
Fruitful discussions with Bing-Nan Lu, Zhen-Hua-Zhang, Yu-Ting Rong and Xiao-Qian Wang are gratefully acknowledged. 
This work is supported by the National Natural Science Foundation of China (Grant Nos. 12347139, 12047503, 12375118, 12435008, and W2412043), the Chinese Academy of Sciences (CAS) President’s International Fellowship Initiative (PIFI, Grant No. 2024PVA0003), the National Key R\&D Program of China (Grant Nos. 2023YFA1606503 and 2024YFE0109801) and the CAS Strategic Priority Research Program (Grant Nos. XDB34010000 and XDB0920000). The results described in this paper are obtained on the High-performance Computing Cluster of ITP-CAS and the ScGrid of the Supercomputing Center, Computer Network Information Center of Chinese Academy of Sciences. 
\end{acknowledgments}
 
% Create the reference section using BibTeX:
 
\bibliography{apssamp.bib}

%apsrev4-2.bst 2019-01-14 (MD) hand-edited version of apsrev4-1.bst
%Control: key (0)
%Control: author (72) initials jnrlst
%Control: editor formatted (1) identically to author
%Control: production of article title (-1) disabled
%Control: page (0) single
%Control: year (1) truncated
%Control: production of eprint (0) enabled
\begin{thebibliography}{93}%
\makeatletter
\providecommand \@ifxundefined [1]{%
 \@ifx{#1\undefined}
}%
\providecommand \@ifnum [1]{%
 \ifnum #1\expandafter \@firstoftwo
 \else \expandafter \@secondoftwo
 \fi
}%
\providecommand \@ifx [1]{%
 \ifx #1\expandafter \@firstoftwo
 \else \expandafter \@secondoftwo
 \fi
}%
\providecommand \natexlab [1]{#1}%
\providecommand \enquote  [1]{``#1''}%
\providecommand \bibnamefont  [1]{#1}%
\providecommand \bibfnamefont [1]{#1}%
\providecommand \citenamefont [1]{#1}%
\providecommand \href@noop [0]{\@secondoftwo}%
\providecommand \href [0]{\begingroup \@sanitize@url \@href}%
\providecommand \@href[1]{\@@startlink{#1}\@@href}%
\providecommand \@@href[1]{\endgroup#1\@@endlink}%
\providecommand \@sanitize@url [0]{\catcode `\\12\catcode `\$12\catcode `\&12\catcode `\#12\catcode `\^12\catcode `\_12\catcode `\%12\relax}%
\providecommand \@@startlink[1]{}%
\providecommand \@@endlink[0]{}%
\providecommand \url  [0]{\begingroup\@sanitize@url \@url }%
\providecommand \@url [1]{\endgroup\@href {#1}{\urlprefix }}%
\providecommand \urlprefix  [0]{URL }%
\providecommand \Eprint [0]{\href }%
\providecommand \doibase [0]{https://doi.org/}%
\providecommand \selectlanguage [0]{\@gobble}%
\providecommand \bibinfo  [0]{\@secondoftwo}%
\providecommand \bibfield  [0]{\@secondoftwo}%
\providecommand \translation [1]{[#1]}%
\providecommand \BibitemOpen [0]{}%
\providecommand \bibitemStop [0]{}%
\providecommand \bibitemNoStop [0]{.\EOS\space}%
\providecommand \EOS [0]{\spacefactor3000\relax}%
\providecommand \BibitemShut  [1]{\csname bibitem#1\endcsname}%
\let\auto@bib@innerbib\@empty
%</preamble>
\bibitem [{\citenamefont {Butler}\ and\ \citenamefont {Nazarewicz}(1996)}]{Butler96}%
  \BibitemOpen
  \bibfield  {author} {\bibinfo {author} {\bibfnamefont {P.~A.}\ \bibnamefont {Butler}}\ and\ \bibinfo {author} {\bibfnamefont {W.}~\bibnamefont {Nazarewicz}},\ }\href {https://doi.org/10.1103/RevModPhys.68.349} {\bibfield  {journal} {\bibinfo  {journal} {Rev. Mod. Phys.}\ }\textbf {\bibinfo {volume} {68}},\ \bibinfo {pages} {349} (\bibinfo {year} {1996})}\BibitemShut {NoStop}%
\bibitem [{\citenamefont {Butler}\ \emph {et~al.}(2020{\natexlab{a}})\citenamefont {Butler}, \citenamefont {Gaffney}, \citenamefont {Spagnoletti}, \citenamefont {Abrahams}, \citenamefont {Bowry}, \citenamefont {Cederk\"all}, \citenamefont {de~Angelis}, \citenamefont {De~Witte}, \citenamefont {Garrett}, \citenamefont {Goldkuhle}, \citenamefont {Henrich}, \citenamefont {Illana}, \citenamefont {Johnston}, \citenamefont {Joss}, \citenamefont {Keatings}, \citenamefont {Kelly}, \citenamefont {Komorowska}, \citenamefont {Konki}, \citenamefont {Kr\"oll}, \citenamefont {Lozano}, \citenamefont {Nara~Singh}, \citenamefont {O'Donnell}, \citenamefont {Ojala}, \citenamefont {Page}, \citenamefont {Pedersen}, \citenamefont {Raison}, \citenamefont {Reiter}, \citenamefont {Rodriguez}, \citenamefont {Rosiak}, \citenamefont {Rothe}, \citenamefont {Scheck}, \citenamefont {Seidlitz}, \citenamefont {Shneidman}, \citenamefont {Siebeck}, \citenamefont {Sinclair}, \citenamefont {Smith}, \citenamefont {Stryjczyk}, \citenamefont
  {Van~Duppen}, \citenamefont {Vinals}, \citenamefont {Virtanen}, \citenamefont {Warr}, \citenamefont {Wrzosek-Lipska},\ and\ \citenamefont {Zieli\ifmmode~\acute{n}\else \'{n}\fi{}ska}}]{Butler20PRL}%
  \BibitemOpen
  \bibfield  {author} {\bibinfo {author} {\bibfnamefont {P.~A.}\ \bibnamefont {Butler}}, \bibinfo {author} {\bibfnamefont {L.~P.}\ \bibnamefont {Gaffney}}, \bibinfo {author} {\bibfnamefont {P.}~\bibnamefont {Spagnoletti}}, \bibinfo {author} {\bibfnamefont {K.}~\bibnamefont {Abrahams}}, \bibinfo {author} {\bibfnamefont {M.}~\bibnamefont {Bowry}}, \bibinfo {author} {\bibfnamefont {J.}~\bibnamefont {Cederk\"all}}, \bibinfo {author} {\bibfnamefont {G.}~\bibnamefont {de~Angelis}}, \bibinfo {author} {\bibfnamefont {H.}~\bibnamefont {De~Witte}}, \bibinfo {author} {\bibfnamefont {P.~E.}\ \bibnamefont {Garrett}}, \bibinfo {author} {\bibfnamefont {A.}~\bibnamefont {Goldkuhle}}, \bibinfo {author} {\bibfnamefont {C.}~\bibnamefont {Henrich}}, \bibinfo {author} {\bibfnamefont {A.}~\bibnamefont {Illana}}, \bibinfo {author} {\bibfnamefont {K.}~\bibnamefont {Johnston}}, \bibinfo {author} {\bibfnamefont {D.~T.}\ \bibnamefont {Joss}}, \bibinfo {author} {\bibfnamefont {J.~M.}\ \bibnamefont {Keatings}}, \bibinfo {author}
  {\bibfnamefont {N.~A.}\ \bibnamefont {Kelly}}, \bibinfo {author} {\bibfnamefont {M.}~\bibnamefont {Komorowska}}, \bibinfo {author} {\bibfnamefont {J.}~\bibnamefont {Konki}}, \bibinfo {author} {\bibfnamefont {T.}~\bibnamefont {Kr\"oll}}, \bibinfo {author} {\bibfnamefont {M.}~\bibnamefont {Lozano}}, \bibinfo {author} {\bibfnamefont {B.~S.}\ \bibnamefont {Nara~Singh}}, \bibinfo {author} {\bibfnamefont {D.}~\bibnamefont {O'Donnell}}, \bibinfo {author} {\bibfnamefont {J.}~\bibnamefont {Ojala}}, \bibinfo {author} {\bibfnamefont {R.~D.}\ \bibnamefont {Page}}, \bibinfo {author} {\bibfnamefont {L.~G.}\ \bibnamefont {Pedersen}}, \bibinfo {author} {\bibfnamefont {C.}~\bibnamefont {Raison}}, \bibinfo {author} {\bibfnamefont {P.}~\bibnamefont {Reiter}}, \bibinfo {author} {\bibfnamefont {J.~A.}\ \bibnamefont {Rodriguez}}, \bibinfo {author} {\bibfnamefont {D.}~\bibnamefont {Rosiak}}, \bibinfo {author} {\bibfnamefont {S.}~\bibnamefont {Rothe}}, \bibinfo {author} {\bibfnamefont {M.}~\bibnamefont {Scheck}}, \bibinfo {author}
  {\bibfnamefont {M.}~\bibnamefont {Seidlitz}}, \bibinfo {author} {\bibfnamefont {T.~M.}\ \bibnamefont {Shneidman}}, \bibinfo {author} {\bibfnamefont {B.}~\bibnamefont {Siebeck}}, \bibinfo {author} {\bibfnamefont {J.}~\bibnamefont {Sinclair}}, \bibinfo {author} {\bibfnamefont {J.~F.}\ \bibnamefont {Smith}}, \bibinfo {author} {\bibfnamefont {M.}~\bibnamefont {Stryjczyk}}, \bibinfo {author} {\bibfnamefont {P.}~\bibnamefont {Van~Duppen}}, \bibinfo {author} {\bibfnamefont {S.}~\bibnamefont {Vinals}}, \bibinfo {author} {\bibfnamefont {V.}~\bibnamefont {Virtanen}}, \bibinfo {author} {\bibfnamefont {N.}~\bibnamefont {Warr}}, \bibinfo {author} {\bibfnamefont {K.}~\bibnamefont {Wrzosek-Lipska}},\ and\ \bibinfo {author} {\bibfnamefont {M.}~\bibnamefont {Zieli\ifmmode~\acute{n}\else \'{n}\fi{}ska}},\ }\href {https://doi.org/10.1103/PhysRevLett.124.042503} {\bibfield  {journal} {\bibinfo  {journal} {Phys. Rev. Lett.}\ }\textbf {\bibinfo {volume} {124}},\ \bibinfo {pages} {042503} (\bibinfo {year}
  {2020}{\natexlab{a}})}\BibitemShut {NoStop}%
\bibitem [{\citenamefont {Butler}(2016)}]{Butler2016}%
  \BibitemOpen
  \bibfield  {author} {\bibinfo {author} {\bibfnamefont {P.~A.}\ \bibnamefont {Butler}},\ }\href {https://doi.org/10.1088/0954-3899/43/7/073002} {\bibfield  {journal} {\bibinfo  {journal} {J. Phys. G: Nucl. Part. Phys.}\ }\textbf {\bibinfo {volume} {43}},\ \bibinfo {pages} {073002} (\bibinfo {year} {2016})}\BibitemShut {NoStop}%
\bibitem [{\citenamefont {Butler}(2020)}]{Butler20}%
  \BibitemOpen
  \bibfield  {author} {\bibinfo {author} {\bibfnamefont {P.~A.}\ \bibnamefont {Butler}},\ }\href {https://doi.org/10.1098/rspa.2020.0202} {\bibfield  {journal} {\bibinfo  {journal} {Proc. R. Soc. A.}\ }\textbf {\bibinfo {volume} {476}},\ \bibinfo {pages} {20200202} (\bibinfo {year} {2020})}\BibitemShut {NoStop}%
\bibitem [{\citenamefont {{Han}}\ \emph {et~al.}(2023)\citenamefont {{Han}}, \citenamefont {{Liu}},\ and\ \citenamefont {{Wang}}}]{han23}%
  \BibitemOpen
  \bibfield  {author} {\bibinfo {author} {\bibfnamefont {X.~C.}\ \bibnamefont {{Han}}}, \bibinfo {author} {\bibfnamefont {C.}~\bibnamefont {{Liu}}},\ and\ \bibinfo {author} {\bibfnamefont {S.~Y.}\ \bibnamefont {{Wang}}},\ }\href {https://doi.org/10.1142/S0218301323400037} {\bibfield  {journal} {\bibinfo  {journal} {Int. J. Mod. Phys. E}\ }\textbf {\bibinfo {volume} {32}},\ \bibinfo {eid} {2340003} (\bibinfo {year} {2023})}\BibitemShut {NoStop}%
\bibitem [{\citenamefont {Butler}(2024)}]{Butler_2024}%
  \BibitemOpen
  \bibfield  {author} {\bibinfo {author} {\bibfnamefont {P.~A.}\ \bibnamefont {Butler}},\ }\href {https://doi.org/10.1088/1402-4896/ad22c9} {\bibfield  {journal} {\bibinfo  {journal} {Phys. Scr.}\ }\textbf {\bibinfo {volume} {99}},\ \bibinfo {pages} {035302} (\bibinfo {year} {2024})}\BibitemShut {NoStop}%
\bibitem [{\citenamefont {Gaffney}\ \emph {et~al.}(2013)\citenamefont {Gaffney}, \citenamefont {Butler}, \citenamefont {Scheck}, \citenamefont {Hayes}, \citenamefont {Wenander}, \citenamefont {Albers}, \citenamefont {Bastin}, \citenamefont {Bauer}, \citenamefont {Blazhev}, \citenamefont {Bönig}, \citenamefont {Bree}, \citenamefont {Cederkäll}, \citenamefont {Chupp}, \citenamefont {Cline}, \citenamefont {Cocolios}, \citenamefont {Davinson}, \citenamefont {De~Witte}, \citenamefont {Diriken}, \citenamefont {Grahn}, \citenamefont {Herzan}, \citenamefont {Huyse}, \citenamefont {Jenkins}, \citenamefont {Joss}, \citenamefont {Kesteloot}, \citenamefont {Konki}, \citenamefont {Kowalczyk}, \citenamefont {Kröll}, \citenamefont {Kwan}, \citenamefont {Lutter}, \citenamefont {Moschner}, \citenamefont {Napiorkowski}, \citenamefont {Pakarinen}, \citenamefont {Pfeiffer}, \citenamefont {Radeck}, \citenamefont {Reiter}, \citenamefont {Reynders}, \citenamefont {Rigby}, \citenamefont {Robledo}, \citenamefont {Rudigier},
  \citenamefont {Sambi}, \citenamefont {Seidlitz}, \citenamefont {Siebeck}, \citenamefont {Stora}, \citenamefont {Thoele}, \citenamefont {Van~Duppen}, \citenamefont {Vermeulen}, \citenamefont {von Schmid}, \citenamefont {Voulot}, \citenamefont {Warr}, \citenamefont {Wimmer}, \citenamefont {Wrzosek-Lipska}, \citenamefont {Wu},\ and\ \citenamefont {Zielinska}}]{RN949}%
  \BibitemOpen
  \bibfield  {author} {\bibinfo {author} {\bibfnamefont {L.~P.}\ \bibnamefont {Gaffney}}, \bibinfo {author} {\bibfnamefont {P.~A.}\ \bibnamefont {Butler}}, \bibinfo {author} {\bibfnamefont {M.}~\bibnamefont {Scheck}}, \bibinfo {author} {\bibfnamefont {A.~B.}\ \bibnamefont {Hayes}}, \bibinfo {author} {\bibfnamefont {F.}~\bibnamefont {Wenander}}, \bibinfo {author} {\bibfnamefont {M.}~\bibnamefont {Albers}}, \bibinfo {author} {\bibfnamefont {B.}~\bibnamefont {Bastin}}, \bibinfo {author} {\bibfnamefont {C.}~\bibnamefont {Bauer}}, \bibinfo {author} {\bibfnamefont {A.}~\bibnamefont {Blazhev}}, \bibinfo {author} {\bibfnamefont {S.}~\bibnamefont {Bönig}}, \bibinfo {author} {\bibfnamefont {N.}~\bibnamefont {Bree}}, \bibinfo {author} {\bibfnamefont {J.}~\bibnamefont {Cederkäll}}, \bibinfo {author} {\bibfnamefont {T.}~\bibnamefont {Chupp}}, \bibinfo {author} {\bibfnamefont {D.}~\bibnamefont {Cline}}, \bibinfo {author} {\bibfnamefont {T.~E.}\ \bibnamefont {Cocolios}}, \bibinfo {author} {\bibfnamefont {T.}~\bibnamefont
  {Davinson}}, \bibinfo {author} {\bibfnamefont {H.}~\bibnamefont {De~Witte}}, \bibinfo {author} {\bibfnamefont {J.}~\bibnamefont {Diriken}}, \bibinfo {author} {\bibfnamefont {T.}~\bibnamefont {Grahn}}, \bibinfo {author} {\bibfnamefont {A.}~\bibnamefont {Herzan}}, \bibinfo {author} {\bibfnamefont {M.}~\bibnamefont {Huyse}}, \bibinfo {author} {\bibfnamefont {D.~G.}\ \bibnamefont {Jenkins}}, \bibinfo {author} {\bibfnamefont {D.~T.}\ \bibnamefont {Joss}}, \bibinfo {author} {\bibfnamefont {N.}~\bibnamefont {Kesteloot}}, \bibinfo {author} {\bibfnamefont {J.}~\bibnamefont {Konki}}, \bibinfo {author} {\bibfnamefont {M.}~\bibnamefont {Kowalczyk}}, \bibinfo {author} {\bibfnamefont {T.}~\bibnamefont {Kröll}}, \bibinfo {author} {\bibfnamefont {E.}~\bibnamefont {Kwan}}, \bibinfo {author} {\bibfnamefont {R.}~\bibnamefont {Lutter}}, \bibinfo {author} {\bibfnamefont {K.}~\bibnamefont {Moschner}}, \bibinfo {author} {\bibfnamefont {P.}~\bibnamefont {Napiorkowski}}, \bibinfo {author} {\bibfnamefont {J.}~\bibnamefont
  {Pakarinen}}, \bibinfo {author} {\bibfnamefont {M.}~\bibnamefont {Pfeiffer}}, \bibinfo {author} {\bibfnamefont {D.}~\bibnamefont {Radeck}}, \bibinfo {author} {\bibfnamefont {P.}~\bibnamefont {Reiter}}, \bibinfo {author} {\bibfnamefont {K.}~\bibnamefont {Reynders}}, \bibinfo {author} {\bibfnamefont {S.~V.}\ \bibnamefont {Rigby}}, \bibinfo {author} {\bibfnamefont {L.~M.}\ \bibnamefont {Robledo}}, \bibinfo {author} {\bibfnamefont {M.}~\bibnamefont {Rudigier}}, \bibinfo {author} {\bibfnamefont {S.}~\bibnamefont {Sambi}}, \bibinfo {author} {\bibfnamefont {M.}~\bibnamefont {Seidlitz}}, \bibinfo {author} {\bibfnamefont {B.}~\bibnamefont {Siebeck}}, \bibinfo {author} {\bibfnamefont {T.}~\bibnamefont {Stora}}, \bibinfo {author} {\bibfnamefont {P.}~\bibnamefont {Thoele}}, \bibinfo {author} {\bibfnamefont {P.}~\bibnamefont {Van~Duppen}}, \bibinfo {author} {\bibfnamefont {M.~J.}\ \bibnamefont {Vermeulen}}, \bibinfo {author} {\bibfnamefont {M.}~\bibnamefont {von Schmid}}, \bibinfo {author} {\bibfnamefont
  {D.}~\bibnamefont {Voulot}}, \bibinfo {author} {\bibfnamefont {N.}~\bibnamefont {Warr}}, \bibinfo {author} {\bibfnamefont {K.}~\bibnamefont {Wimmer}}, \bibinfo {author} {\bibfnamefont {K.}~\bibnamefont {Wrzosek-Lipska}}, \bibinfo {author} {\bibfnamefont {C.~Y.}\ \bibnamefont {Wu}},\ and\ \bibinfo {author} {\bibfnamefont {M.}~\bibnamefont {Zielinska}},\ }\href {https://doi.org/10.1038/nature12073} {\bibfield  {journal} {\bibinfo  {journal} {Nature}\ }\textbf {\bibinfo {volume} {497}},\ \bibinfo {pages} {199} (\bibinfo {year} {2013})}\BibitemShut {NoStop}%
\bibitem [{\citenamefont {Butler}\ \emph {et~al.}(2020{\natexlab{b}})\citenamefont {Butler}, \citenamefont {Gaffney}, \citenamefont {Spagnoletti}, \citenamefont {Abrahams}, \citenamefont {Bowry}, \citenamefont {Cederk\"all}, \citenamefont {de~Angelis}, \citenamefont {De~Witte}, \citenamefont {Garrett}, \citenamefont {Goldkuhle}, \citenamefont {Henrich}, \citenamefont {Illana}, \citenamefont {Johnston}, \citenamefont {Joss}, \citenamefont {Keatings}, \citenamefont {Kelly}, \citenamefont {Komorowska}, \citenamefont {Konki}, \citenamefont {Kr\"oll}, \citenamefont {Lozano}, \citenamefont {Nara~Singh}, \citenamefont {O'Donnell}, \citenamefont {Ojala}, \citenamefont {Page}, \citenamefont {Pedersen}, \citenamefont {Raison}, \citenamefont {Reiter}, \citenamefont {Rodriguez}, \citenamefont {Rosiak}, \citenamefont {Rothe}, \citenamefont {Scheck}, \citenamefont {Seidlitz}, \citenamefont {Shneidman}, \citenamefont {Siebeck}, \citenamefont {Sinclair}, \citenamefont {Smith}, \citenamefont {Stryjczyk}, \citenamefont
  {Van~Duppen}, \citenamefont {Vinals}, \citenamefont {Virtanen}, \citenamefont {Warr}, \citenamefont {Wrzosek-Lipska},\ and\ \citenamefont {Zieli\ifmmode~\acute{n}\else \'{n}\fi{}ska}}]{224Ra20}%
  \BibitemOpen
  \bibfield  {author} {\bibinfo {author} {\bibfnamefont {P.~A.}\ \bibnamefont {Butler}}, \bibinfo {author} {\bibfnamefont {L.~P.}\ \bibnamefont {Gaffney}}, \bibinfo {author} {\bibfnamefont {P.}~\bibnamefont {Spagnoletti}}, \bibinfo {author} {\bibfnamefont {K.}~\bibnamefont {Abrahams}}, \bibinfo {author} {\bibfnamefont {M.}~\bibnamefont {Bowry}}, \bibinfo {author} {\bibfnamefont {J.}~\bibnamefont {Cederk\"all}}, \bibinfo {author} {\bibfnamefont {G.}~\bibnamefont {de~Angelis}}, \bibinfo {author} {\bibfnamefont {H.}~\bibnamefont {De~Witte}}, \bibinfo {author} {\bibfnamefont {P.~E.}\ \bibnamefont {Garrett}}, \bibinfo {author} {\bibfnamefont {A.}~\bibnamefont {Goldkuhle}}, \bibinfo {author} {\bibfnamefont {C.}~\bibnamefont {Henrich}}, \bibinfo {author} {\bibfnamefont {A.}~\bibnamefont {Illana}}, \bibinfo {author} {\bibfnamefont {K.}~\bibnamefont {Johnston}}, \bibinfo {author} {\bibfnamefont {D.~T.}\ \bibnamefont {Joss}}, \bibinfo {author} {\bibfnamefont {J.~M.}\ \bibnamefont {Keatings}}, \bibinfo {author}
  {\bibfnamefont {N.~A.}\ \bibnamefont {Kelly}}, \bibinfo {author} {\bibfnamefont {M.}~\bibnamefont {Komorowska}}, \bibinfo {author} {\bibfnamefont {J.}~\bibnamefont {Konki}}, \bibinfo {author} {\bibfnamefont {T.}~\bibnamefont {Kr\"oll}}, \bibinfo {author} {\bibfnamefont {M.}~\bibnamefont {Lozano}}, \bibinfo {author} {\bibfnamefont {B.~S.}\ \bibnamefont {Nara~Singh}}, \bibinfo {author} {\bibfnamefont {D.}~\bibnamefont {O'Donnell}}, \bibinfo {author} {\bibfnamefont {J.}~\bibnamefont {Ojala}}, \bibinfo {author} {\bibfnamefont {R.~D.}\ \bibnamefont {Page}}, \bibinfo {author} {\bibfnamefont {L.~G.}\ \bibnamefont {Pedersen}}, \bibinfo {author} {\bibfnamefont {C.}~\bibnamefont {Raison}}, \bibinfo {author} {\bibfnamefont {P.}~\bibnamefont {Reiter}}, \bibinfo {author} {\bibfnamefont {J.~A.}\ \bibnamefont {Rodriguez}}, \bibinfo {author} {\bibfnamefont {D.}~\bibnamefont {Rosiak}}, \bibinfo {author} {\bibfnamefont {S.}~\bibnamefont {Rothe}}, \bibinfo {author} {\bibfnamefont {M.}~\bibnamefont {Scheck}}, \bibinfo {author}
  {\bibfnamefont {M.}~\bibnamefont {Seidlitz}}, \bibinfo {author} {\bibfnamefont {T.~M.}\ \bibnamefont {Shneidman}}, \bibinfo {author} {\bibfnamefont {B.}~\bibnamefont {Siebeck}}, \bibinfo {author} {\bibfnamefont {J.}~\bibnamefont {Sinclair}}, \bibinfo {author} {\bibfnamefont {J.~F.}\ \bibnamefont {Smith}}, \bibinfo {author} {\bibfnamefont {M.}~\bibnamefont {Stryjczyk}}, \bibinfo {author} {\bibfnamefont {P.}~\bibnamefont {Van~Duppen}}, \bibinfo {author} {\bibfnamefont {S.}~\bibnamefont {Vinals}}, \bibinfo {author} {\bibfnamefont {V.}~\bibnamefont {Virtanen}}, \bibinfo {author} {\bibfnamefont {N.}~\bibnamefont {Warr}}, \bibinfo {author} {\bibfnamefont {K.}~\bibnamefont {Wrzosek-Lipska}},\ and\ \bibinfo {author} {\bibfnamefont {M.}~\bibnamefont {Zieli\ifmmode~\acute{n}\else \'{n}\fi{}ska}},\ }\href {https://doi.org/10.1103/PhysRevLett.124.042503} {\bibfield  {journal} {\bibinfo  {journal} {Phys. Rev. Lett.}\ }\textbf {\bibinfo {volume} {124}},\ \bibinfo {pages} {042503} (\bibinfo {year}
  {2020}{\natexlab{b}})}\BibitemShut {NoStop}%
\bibitem [{\citenamefont {Chishti}\ \emph {et~al.}(2020)\citenamefont {Chishti}, \citenamefont {O'Donnell}, \citenamefont {Battaglia}, \citenamefont {Bowry}, \citenamefont {Jaroszynski}, \citenamefont {Singh}, \citenamefont {Scheck}, \citenamefont {Spagnoletti},\ and\ \citenamefont {Smith}}]{RN950}%
  \BibitemOpen
  \bibfield  {author} {\bibinfo {author} {\bibfnamefont {M.~M.~R.}\ \bibnamefont {Chishti}}, \bibinfo {author} {\bibfnamefont {D.}~\bibnamefont {O'Donnell}}, \bibinfo {author} {\bibfnamefont {G.}~\bibnamefont {Battaglia}}, \bibinfo {author} {\bibfnamefont {M.}~\bibnamefont {Bowry}}, \bibinfo {author} {\bibfnamefont {D.~A.}\ \bibnamefont {Jaroszynski}}, \bibinfo {author} {\bibfnamefont {B.~S.~N.}\ \bibnamefont {Singh}}, \bibinfo {author} {\bibfnamefont {M.}~\bibnamefont {Scheck}}, \bibinfo {author} {\bibfnamefont {P.}~\bibnamefont {Spagnoletti}},\ and\ \bibinfo {author} {\bibfnamefont {J.~F.}\ \bibnamefont {Smith}},\ }\href {https://doi.org/10.1038/s41567-020-0899-4} {\bibfield  {journal} {\bibinfo  {journal} {Nat. Phys.}\ }\textbf {\bibinfo {volume} {16}},\ \bibinfo {pages} {853} (\bibinfo {year} {2020})}\BibitemShut {NoStop}%
\bibitem [{\citenamefont {Bucher}\ \emph {et~al.}(2016)\citenamefont {Bucher}, \citenamefont {Zhu}, \citenamefont {Wu}, \citenamefont {Janssens}, \citenamefont {Cline}, \citenamefont {Hayes}, \citenamefont {Albers}, \citenamefont {Ayangeakaa}, \citenamefont {Butler}, \citenamefont {Campbell}, \citenamefont {Carpenter}, \citenamefont {Chiara}, \citenamefont {Clark}, \citenamefont {Crawford}, \citenamefont {Cromaz}, \citenamefont {David}, \citenamefont {Dickerson}, \citenamefont {Gregor}, \citenamefont {Harker}, \citenamefont {Hoffman}, \citenamefont {Kay}, \citenamefont {Kondev}, \citenamefont {Korichi}, \citenamefont {Lauritsen}, \citenamefont {Macchiavelli}, \citenamefont {Pardo}, \citenamefont {Richard}, \citenamefont {Riley}, \citenamefont {Savard}, \citenamefont {Scheck}, \citenamefont {Seweryniak}, \citenamefont {Smith}, \citenamefont {Vondrasek},\ and\ \citenamefont {Wiens}}]{144Ba16}%
  \BibitemOpen
  \bibfield  {author} {\bibinfo {author} {\bibfnamefont {B.}~\bibnamefont {Bucher}}, \bibinfo {author} {\bibfnamefont {S.}~\bibnamefont {Zhu}}, \bibinfo {author} {\bibfnamefont {C.~Y.}\ \bibnamefont {Wu}}, \bibinfo {author} {\bibfnamefont {R.~V.~F.}\ \bibnamefont {Janssens}}, \bibinfo {author} {\bibfnamefont {D.}~\bibnamefont {Cline}}, \bibinfo {author} {\bibfnamefont {A.~B.}\ \bibnamefont {Hayes}}, \bibinfo {author} {\bibfnamefont {M.}~\bibnamefont {Albers}}, \bibinfo {author} {\bibfnamefont {A.~D.}\ \bibnamefont {Ayangeakaa}}, \bibinfo {author} {\bibfnamefont {P.~A.}\ \bibnamefont {Butler}}, \bibinfo {author} {\bibfnamefont {C.~M.}\ \bibnamefont {Campbell}}, \bibinfo {author} {\bibfnamefont {M.~P.}\ \bibnamefont {Carpenter}}, \bibinfo {author} {\bibfnamefont {C.~J.}\ \bibnamefont {Chiara}}, \bibinfo {author} {\bibfnamefont {J.~A.}\ \bibnamefont {Clark}}, \bibinfo {author} {\bibfnamefont {H.~L.}\ \bibnamefont {Crawford}}, \bibinfo {author} {\bibfnamefont {M.}~\bibnamefont {Cromaz}}, \bibinfo {author}
  {\bibfnamefont {H.~M.}\ \bibnamefont {David}}, \bibinfo {author} {\bibfnamefont {C.}~\bibnamefont {Dickerson}}, \bibinfo {author} {\bibfnamefont {E.~T.}\ \bibnamefont {Gregor}}, \bibinfo {author} {\bibfnamefont {J.}~\bibnamefont {Harker}}, \bibinfo {author} {\bibfnamefont {C.~R.}\ \bibnamefont {Hoffman}}, \bibinfo {author} {\bibfnamefont {B.~P.}\ \bibnamefont {Kay}}, \bibinfo {author} {\bibfnamefont {F.~G.}\ \bibnamefont {Kondev}}, \bibinfo {author} {\bibfnamefont {A.}~\bibnamefont {Korichi}}, \bibinfo {author} {\bibfnamefont {T.}~\bibnamefont {Lauritsen}}, \bibinfo {author} {\bibfnamefont {A.~O.}\ \bibnamefont {Macchiavelli}}, \bibinfo {author} {\bibfnamefont {R.~C.}\ \bibnamefont {Pardo}}, \bibinfo {author} {\bibfnamefont {A.}~\bibnamefont {Richard}}, \bibinfo {author} {\bibfnamefont {M.~A.}\ \bibnamefont {Riley}}, \bibinfo {author} {\bibfnamefont {G.}~\bibnamefont {Savard}}, \bibinfo {author} {\bibfnamefont {M.}~\bibnamefont {Scheck}}, \bibinfo {author} {\bibfnamefont {D.}~\bibnamefont {Seweryniak}},
  \bibinfo {author} {\bibfnamefont {M.~K.}\ \bibnamefont {Smith}}, \bibinfo {author} {\bibfnamefont {R.}~\bibnamefont {Vondrasek}},\ and\ \bibinfo {author} {\bibfnamefont {A.}~\bibnamefont {Wiens}},\ }\href {https://doi.org/10.1103/PhysRevLett.116.112503} {\bibfield  {journal} {\bibinfo  {journal} {Phys. Rev. Lett.}\ }\textbf {\bibinfo {volume} {116}},\ \bibinfo {pages} {112503} (\bibinfo {year} {2016})}\BibitemShut {NoStop}%
\bibitem [{\citenamefont {Bucher}\ \emph {et~al.}(2017)\citenamefont {Bucher}, \citenamefont {Zhu}, \citenamefont {Wu}, \citenamefont {Janssens}, \citenamefont {Bernard}, \citenamefont {Robledo}, \citenamefont {Rodr\'{\i}guez}, \citenamefont {Cline}, \citenamefont {Hayes}, \citenamefont {Ayangeakaa}, \citenamefont {Buckner}, \citenamefont {Campbell}, \citenamefont {Carpenter}, \citenamefont {Clark}, \citenamefont {Crawford}, \citenamefont {David}, \citenamefont {Dickerson}, \citenamefont {Harker}, \citenamefont {Hoffman}, \citenamefont {Kay}, \citenamefont {Kondev}, \citenamefont {Lauritsen}, \citenamefont {Macchiavelli}, \citenamefont {Pardo}, \citenamefont {Savard}, \citenamefont {Seweryniak},\ and\ \citenamefont {Vondrasek}}]{146Ba17}%
  \BibitemOpen
  \bibfield  {author} {\bibinfo {author} {\bibfnamefont {B.}~\bibnamefont {Bucher}}, \bibinfo {author} {\bibfnamefont {S.}~\bibnamefont {Zhu}}, \bibinfo {author} {\bibfnamefont {C.~Y.}\ \bibnamefont {Wu}}, \bibinfo {author} {\bibfnamefont {R.~V.~F.}\ \bibnamefont {Janssens}}, \bibinfo {author} {\bibfnamefont {R.~N.}\ \bibnamefont {Bernard}}, \bibinfo {author} {\bibfnamefont {L.~M.}\ \bibnamefont {Robledo}}, \bibinfo {author} {\bibfnamefont {T.~R.}\ \bibnamefont {Rodr\'{\i}guez}}, \bibinfo {author} {\bibfnamefont {D.}~\bibnamefont {Cline}}, \bibinfo {author} {\bibfnamefont {A.~B.}\ \bibnamefont {Hayes}}, \bibinfo {author} {\bibfnamefont {A.~D.}\ \bibnamefont {Ayangeakaa}}, \bibinfo {author} {\bibfnamefont {M.~Q.}\ \bibnamefont {Buckner}}, \bibinfo {author} {\bibfnamefont {C.~M.}\ \bibnamefont {Campbell}}, \bibinfo {author} {\bibfnamefont {M.~P.}\ \bibnamefont {Carpenter}}, \bibinfo {author} {\bibfnamefont {J.~A.}\ \bibnamefont {Clark}}, \bibinfo {author} {\bibfnamefont {H.~L.}\ \bibnamefont {Crawford}},
  \bibinfo {author} {\bibfnamefont {H.~M.}\ \bibnamefont {David}}, \bibinfo {author} {\bibfnamefont {C.}~\bibnamefont {Dickerson}}, \bibinfo {author} {\bibfnamefont {J.}~\bibnamefont {Harker}}, \bibinfo {author} {\bibfnamefont {C.~R.}\ \bibnamefont {Hoffman}}, \bibinfo {author} {\bibfnamefont {B.~P.}\ \bibnamefont {Kay}}, \bibinfo {author} {\bibfnamefont {F.~G.}\ \bibnamefont {Kondev}}, \bibinfo {author} {\bibfnamefont {T.}~\bibnamefont {Lauritsen}}, \bibinfo {author} {\bibfnamefont {A.~O.}\ \bibnamefont {Macchiavelli}}, \bibinfo {author} {\bibfnamefont {R.~C.}\ \bibnamefont {Pardo}}, \bibinfo {author} {\bibfnamefont {G.}~\bibnamefont {Savard}}, \bibinfo {author} {\bibfnamefont {D.}~\bibnamefont {Seweryniak}},\ and\ \bibinfo {author} {\bibfnamefont {R.}~\bibnamefont {Vondrasek}},\ }\href {https://doi.org/10.1103/PhysRevLett.118.152504} {\bibfield  {journal} {\bibinfo  {journal} {Phys. Rev. Lett.}\ }\textbf {\bibinfo {volume} {118}},\ \bibinfo {pages} {152504} (\bibinfo {year} {2017})}\BibitemShut {NoStop}%
\bibitem [{\citenamefont {.}\ \emph {et~al.}(2023)\citenamefont {.}, \citenamefont {Kumar}, \citenamefont {Kumar}, \citenamefont {Rajput}, \citenamefont {Devi}, \citenamefont {Kumar}, \citenamefont {Ahmad}, \citenamefont {Banerjee}, \citenamefont {Rohilla}, \citenamefont {Gupta}, \citenamefont {Pandey}, \citenamefont {Bhushan}, \citenamefont {Gupta}, \citenamefont {Mandal}, \citenamefont {Bhattacharjee}, \citenamefont {Arora}, \citenamefont {Garg}, \citenamefont {Bala}, \citenamefont {Kumar}, \citenamefont {Singh}, \citenamefont {Muralithar}, \citenamefont {Jyothi}, \citenamefont {Majumder}, \citenamefont {Sharma}, \citenamefont {Sharma},\ and\ \citenamefont {Dhiman}}]{2023IndicationsOO}%
  \BibitemOpen
  \bibfield  {author} {\bibinfo {author} {\bibfnamefont {A.}~\bibnamefont {.}}, \bibinfo {author} {\bibfnamefont {S.}~\bibnamefont {Kumar}}, \bibinfo {author} {\bibfnamefont {N.}~\bibnamefont {Kumar}}, \bibinfo {author} {\bibfnamefont {N.}~\bibnamefont {Rajput}}, \bibinfo {author} {\bibfnamefont {K.~R.}\ \bibnamefont {Devi}}, \bibinfo {author} {\bibfnamefont {N.}~\bibnamefont {Kumar}}, \bibinfo {author} {\bibfnamefont {C.~V.}\ \bibnamefont {Ahmad}}, \bibinfo {author} {\bibfnamefont {A.}~\bibnamefont {Banerjee}}, \bibinfo {author} {\bibfnamefont {A.}~\bibnamefont {Rohilla}}, \bibinfo {author} {\bibfnamefont {C.}~\bibnamefont {Gupta}}, \bibinfo {author} {\bibfnamefont {A.}~\bibnamefont {Pandey}}, \bibinfo {author} {\bibfnamefont {R.}~\bibnamefont {Bhushan}}, \bibinfo {author} {\bibfnamefont {U.}~\bibnamefont {Gupta}}, \bibinfo {author} {\bibfnamefont {S.}~\bibnamefont {Mandal}}, \bibinfo {author} {\bibfnamefont {S.~S.}\ \bibnamefont {Bhattacharjee}}, \bibinfo {author} {\bibfnamefont {D.}~\bibnamefont {Arora}},
  \bibinfo {author} {\bibfnamefont {R.}~\bibnamefont {Garg}}, \bibinfo {author} {\bibfnamefont {I.}~\bibnamefont {Bala}}, \bibinfo {author} {\bibfnamefont {R.}~\bibnamefont {Kumar}}, \bibinfo {author} {\bibfnamefont {R.~P.}\ \bibnamefont {Singh}}, \bibinfo {author} {\bibfnamefont {S.}~\bibnamefont {Muralithar}}, \bibinfo {author} {\bibfnamefont {V.~V.}\ \bibnamefont {Jyothi}}, \bibinfo {author} {\bibfnamefont {C.}~\bibnamefont {Majumder}}, \bibinfo {author} {\bibfnamefont {H.}~\bibnamefont {Sharma}}, \bibinfo {author} {\bibfnamefont {A.}~\bibnamefont {Sharma}},\ and\ \bibinfo {author} {\bibfnamefont {S.}~\bibnamefont {Dhiman}},\ }\href {https://api.semanticscholar.org/CorpusID:260034300} {\bibfield  {journal} {\bibinfo  {journal} {Phys. Scr.}\ }\textbf {\bibinfo {volume} {98}} (\bibinfo {year} {2023})}\BibitemShut {NoStop}%
\bibitem [{\citenamefont {Wang}\ \emph {et~al.}(2022{\natexlab{a}})\citenamefont {Wang}, \citenamefont {Han}, \citenamefont {Xu}, \citenamefont {Hua}, \citenamefont {Bark}, \citenamefont {Zhang}, \citenamefont {Wang}, \citenamefont {Shneidman}, \citenamefont {Zhou}, \citenamefont {Meng}, \citenamefont {Wyngaardt}, \citenamefont {Dai}, \citenamefont {Xu}, \citenamefont {Li}, \citenamefont {Li}, \citenamefont {Ye}, \citenamefont {Jiang}, \citenamefont {Li}, \citenamefont {Niu}, \citenamefont {Chen}, \citenamefont {Wu}, \citenamefont {Luo}, \citenamefont {Wang}, \citenamefont {Sun}, \citenamefont {Liu}, \citenamefont {Li}, \citenamefont {Zhang}, \citenamefont {Guo}, \citenamefont {Jones}, \citenamefont {Lawrie}, \citenamefont {Lawrie}, \citenamefont {Sharpey-Schafer}, \citenamefont {Wiedeking}, \citenamefont {Majola}, \citenamefont {Bucher}, \citenamefont {Dinoko}, \citenamefont {Maqabuka}, \citenamefont {Makhathini}, \citenamefont {Mdletshe}, \citenamefont {Shirinda},\ and\ \citenamefont {Sowazi}}]{ge712022}%
  \BibitemOpen
  \bibfield  {author} {\bibinfo {author} {\bibfnamefont {C.~G.}\ \bibnamefont {Wang}}, \bibinfo {author} {\bibfnamefont {R.}~\bibnamefont {Han}}, \bibinfo {author} {\bibfnamefont {C.}~\bibnamefont {Xu}}, \bibinfo {author} {\bibfnamefont {H.}~\bibnamefont {Hua}}, \bibinfo {author} {\bibfnamefont {R.~A.}\ \bibnamefont {Bark}}, \bibinfo {author} {\bibfnamefont {S.~Q.}\ \bibnamefont {Zhang}}, \bibinfo {author} {\bibfnamefont {S.~Y.}\ \bibnamefont {Wang}}, \bibinfo {author} {\bibfnamefont {T.~M.}\ \bibnamefont {Shneidman}}, \bibinfo {author} {\bibfnamefont {S.~G.}\ \bibnamefont {Zhou}}, \bibinfo {author} {\bibfnamefont {J.}~\bibnamefont {Meng}}, \bibinfo {author} {\bibfnamefont {S.~M.}\ \bibnamefont {Wyngaardt}}, \bibinfo {author} {\bibfnamefont {A.~C.}\ \bibnamefont {Dai}}, \bibinfo {author} {\bibfnamefont {F.~R.}\ \bibnamefont {Xu}}, \bibinfo {author} {\bibfnamefont {X.~Q.}\ \bibnamefont {Li}}, \bibinfo {author} {\bibfnamefont {Z.~H.}\ \bibnamefont {Li}}, \bibinfo {author} {\bibfnamefont {Y.~L.}\ \bibnamefont
  {Ye}}, \bibinfo {author} {\bibfnamefont {D.~X.}\ \bibnamefont {Jiang}}, \bibinfo {author} {\bibfnamefont {C.~G.}\ \bibnamefont {Li}}, \bibinfo {author} {\bibfnamefont {C.~Y.}\ \bibnamefont {Niu}}, \bibinfo {author} {\bibfnamefont {Z.~Q.}\ \bibnamefont {Chen}}, \bibinfo {author} {\bibfnamefont {H.~Y.}\ \bibnamefont {Wu}}, \bibinfo {author} {\bibfnamefont {D.~W.}\ \bibnamefont {Luo}}, \bibinfo {author} {\bibfnamefont {S.}~\bibnamefont {Wang}}, \bibinfo {author} {\bibfnamefont {D.~P.}\ \bibnamefont {Sun}}, \bibinfo {author} {\bibfnamefont {C.}~\bibnamefont {Liu}}, \bibinfo {author} {\bibfnamefont {Z.~Q.}\ \bibnamefont {Li}}, \bibinfo {author} {\bibfnamefont {N.~B.}\ \bibnamefont {Zhang}}, \bibinfo {author} {\bibfnamefont {R.~J.}\ \bibnamefont {Guo}}, \bibinfo {author} {\bibfnamefont {P.}~\bibnamefont {Jones}}, \bibinfo {author} {\bibfnamefont {E.~A.}\ \bibnamefont {Lawrie}}, \bibinfo {author} {\bibfnamefont {J.~J.}\ \bibnamefont {Lawrie}}, \bibinfo {author} {\bibfnamefont {J.~F.}\ \bibnamefont
  {Sharpey-Schafer}}, \bibinfo {author} {\bibfnamefont {M.}~\bibnamefont {Wiedeking}}, \bibinfo {author} {\bibfnamefont {S.~N.~T.}\ \bibnamefont {Majola}}, \bibinfo {author} {\bibfnamefont {T.~D.}\ \bibnamefont {Bucher}}, \bibinfo {author} {\bibfnamefont {T.}~\bibnamefont {Dinoko}}, \bibinfo {author} {\bibfnamefont {B.}~\bibnamefont {Maqabuka}}, \bibinfo {author} {\bibfnamefont {L.}~\bibnamefont {Makhathini}}, \bibinfo {author} {\bibfnamefont {L.}~\bibnamefont {Mdletshe}}, \bibinfo {author} {\bibfnamefont {O.}~\bibnamefont {Shirinda}},\ and\ \bibinfo {author} {\bibfnamefont {K.}~\bibnamefont {Sowazi}},\ }\href {https://doi.org/10.1103/PhysRevC.106.L011303} {\bibfield  {journal} {\bibinfo  {journal} {Phys. Rev. C}\ }\textbf {\bibinfo {volume} {106}},\ \bibinfo {pages} {L011303} (\bibinfo {year} {2022}{\natexlab{a}})}\BibitemShut {NoStop}%
\bibitem [{\citenamefont {Zhou}\ \emph {et~al.}(2024{\natexlab{a}})\citenamefont {Zhou}, \citenamefont {Luo}, \citenamefont {Wu}, \citenamefont {Wang}, \citenamefont {Niu}, \citenamefont {Zhang}, \citenamefont {Xu}, \citenamefont {Li}, \citenamefont {Li}, \citenamefont {Hua}, \citenamefont {Zhang}, \citenamefont {Guo}, \citenamefont {Li}, \citenamefont {Qiang}, \citenamefont {Rohilla}, \citenamefont {Lin}, \citenamefont {Zhang}, \citenamefont {Ni}, \citenamefont {Zhang}, \citenamefont {Wang}, \citenamefont {Fang}, \citenamefont {Raju}, \citenamefont {Bing}, \citenamefont {Zhang}, \citenamefont {Huang}, \citenamefont {Liu}, \citenamefont {Zeng}, \citenamefont {Guo},\ and\ \citenamefont {Zhou}}]{ga672024}%
  \BibitemOpen
  \bibfield  {author} {\bibinfo {author} {\bibfnamefont {Z.~X.}\ \bibnamefont {Zhou}}, \bibinfo {author} {\bibfnamefont {D.~W.}\ \bibnamefont {Luo}}, \bibinfo {author} {\bibfnamefont {H.~Y.}\ \bibnamefont {Wu}}, \bibinfo {author} {\bibfnamefont {Y.~Y.}\ \bibnamefont {Wang}}, \bibinfo {author} {\bibfnamefont {Y.~F.}\ \bibnamefont {Niu}}, \bibinfo {author} {\bibfnamefont {W.}~\bibnamefont {Zhang}}, \bibinfo {author} {\bibfnamefont {C.}~\bibnamefont {Xu}}, \bibinfo {author} {\bibfnamefont {G.~S.}\ \bibnamefont {Li}}, \bibinfo {author} {\bibfnamefont {Z.~H.}\ \bibnamefont {Li}}, \bibinfo {author} {\bibfnamefont {H.}~\bibnamefont {Hua}}, \bibinfo {author} {\bibfnamefont {S.~Q.}\ \bibnamefont {Zhang}}, \bibinfo {author} {\bibfnamefont {C.~Y.}\ \bibnamefont {Guo}}, \bibinfo {author} {\bibfnamefont {X.~Q.}\ \bibnamefont {Li}}, \bibinfo {author} {\bibfnamefont {Y.~H.}\ \bibnamefont {Qiang}}, \bibinfo {author} {\bibfnamefont {A.}~\bibnamefont {Rohilla}}, \bibinfo {author} {\bibfnamefont {J.}~\bibnamefont {Lin}},
  \bibinfo {author} {\bibfnamefont {J.~Z.}\ \bibnamefont {Zhang}}, \bibinfo {author} {\bibfnamefont {L.}~\bibnamefont {Ni}}, \bibinfo {author} {\bibfnamefont {S.~Y.}\ \bibnamefont {Zhang}}, \bibinfo {author} {\bibfnamefont {J.~G.}\ \bibnamefont {Wang}}, \bibinfo {author} {\bibfnamefont {Y.~D.}\ \bibnamefont {Fang}}, \bibinfo {author} {\bibfnamefont {M.~K.}\ \bibnamefont {Raju}}, \bibinfo {author} {\bibfnamefont {D.}~\bibnamefont {Bing}}, \bibinfo {author} {\bibfnamefont {W.~Q.}\ \bibnamefont {Zhang}}, \bibinfo {author} {\bibfnamefont {H.}~\bibnamefont {Huang}}, \bibinfo {author} {\bibfnamefont {M.~L.}\ \bibnamefont {Liu}}, \bibinfo {author} {\bibfnamefont {F.~F.}\ \bibnamefont {Zeng}}, \bibinfo {author} {\bibfnamefont {S.}~\bibnamefont {Guo}},\ and\ \bibinfo {author} {\bibfnamefont {X.~H.}\ \bibnamefont {Zhou}},\ }\href {https://doi.org/10.1103/PhysRevC.110.024309} {\bibfield  {journal} {\bibinfo  {journal} {Phys. Rev. C}\ }\textbf {\bibinfo {volume} {110}},\ \bibinfo {pages} {024309} (\bibinfo {year}
  {2024}{\natexlab{a}})}\BibitemShut {NoStop}%
\bibitem [{\citenamefont {Rudolph}\ \emph {et~al.}(1999)\citenamefont {Rudolph}, \citenamefont {Baktash}, \citenamefont {Brinkman}, \citenamefont {Caurier}, \citenamefont {Dean}, \citenamefont {Devlin}, \citenamefont {Dobaczewski}, \citenamefont {Heenen}, \citenamefont {Jin}, \citenamefont {LaFosse}, \citenamefont {Nazarewicz}, \citenamefont {Nowacki}, \citenamefont {Poves}, \citenamefont {Riedinger}, \citenamefont {Sarantites}, \citenamefont {Satuła},\ and\ \citenamefont {Yu}}]{RN619}%
  \BibitemOpen
  \bibfield  {author} {\bibinfo {author} {\bibfnamefont {D.}~\bibnamefont {Rudolph}}, \bibinfo {author} {\bibfnamefont {C.}~\bibnamefont {Baktash}}, \bibinfo {author} {\bibfnamefont {M.~J.}\ \bibnamefont {Brinkman}}, \bibinfo {author} {\bibfnamefont {E.}~\bibnamefont {Caurier}}, \bibinfo {author} {\bibfnamefont {D.~J.}\ \bibnamefont {Dean}}, \bibinfo {author} {\bibfnamefont {M.}~\bibnamefont {Devlin}}, \bibinfo {author} {\bibfnamefont {J.}~\bibnamefont {Dobaczewski}}, \bibinfo {author} {\bibfnamefont {P.~H.}\ \bibnamefont {Heenen}}, \bibinfo {author} {\bibfnamefont {H.~Q.}\ \bibnamefont {Jin}}, \bibinfo {author} {\bibfnamefont {D.~R.}\ \bibnamefont {LaFosse}}, \bibinfo {author} {\bibfnamefont {W.}~\bibnamefont {Nazarewicz}}, \bibinfo {author} {\bibfnamefont {F.}~\bibnamefont {Nowacki}}, \bibinfo {author} {\bibfnamefont {A.}~\bibnamefont {Poves}}, \bibinfo {author} {\bibfnamefont {L.~L.}\ \bibnamefont {Riedinger}}, \bibinfo {author} {\bibfnamefont {D.~G.}\ \bibnamefont {Sarantites}}, \bibinfo {author}
  {\bibfnamefont {W.}~\bibnamefont {Satuła}},\ and\ \bibinfo {author} {\bibfnamefont {C.~H.}\ \bibnamefont {Yu}},\ }\href {https://doi.org/10.1103/PhysRevLett.82.3763} {\bibfield  {journal} {\bibinfo  {journal} {Phys. Rev. Lett.}\ }\textbf {\bibinfo {volume} {82}},\ \bibinfo {pages} {3763} (\bibinfo {year} {1999})}\BibitemShut {NoStop}%
\bibitem [{\citenamefont {Johansson}\ \emph {et~al.}(2006)\citenamefont {Johansson}, \citenamefont {Rudolph}, \citenamefont {Ekman}, \citenamefont {Fahlander}, \citenamefont {Andreoiu}, \citenamefont {Bentley}, \citenamefont {Carpenter}, \citenamefont {Charity}, \citenamefont {Clark}, \citenamefont {Fallon}, \citenamefont {Janssens}, \citenamefont {Kondev}, \citenamefont {Khoo}, \citenamefont {Lauritsen}, \citenamefont {Macchiavelli}, \citenamefont {Reviol}, \citenamefont {Sarantites}, \citenamefont {Seweryniak}, \citenamefont {Svensson},\ and\ \citenamefont {Williams}}]{RN616}%
  \BibitemOpen
  \bibfield  {author} {\bibinfo {author} {\bibfnamefont {E.~K.}\ \bibnamefont {Johansson}}, \bibinfo {author} {\bibfnamefont {D.}~\bibnamefont {Rudolph}}, \bibinfo {author} {\bibfnamefont {J.}~\bibnamefont {Ekman}}, \bibinfo {author} {\bibfnamefont {C.}~\bibnamefont {Fahlander}}, \bibinfo {author} {\bibfnamefont {C.}~\bibnamefont {Andreoiu}}, \bibinfo {author} {\bibfnamefont {M.~A.}\ \bibnamefont {Bentley}}, \bibinfo {author} {\bibfnamefont {M.~P.}\ \bibnamefont {Carpenter}}, \bibinfo {author} {\bibfnamefont {R.~J.}\ \bibnamefont {Charity}}, \bibinfo {author} {\bibfnamefont {R.~M.}\ \bibnamefont {Clark}}, \bibinfo {author} {\bibfnamefont {P.}~\bibnamefont {Fallon}}, \bibinfo {author} {\bibfnamefont {R.~V.~F.}\ \bibnamefont {Janssens}}, \bibinfo {author} {\bibfnamefont {F.~G.}\ \bibnamefont {Kondev}}, \bibinfo {author} {\bibfnamefont {T.~L.}\ \bibnamefont {Khoo}}, \bibinfo {author} {\bibfnamefont {T.}~\bibnamefont {Lauritsen}}, \bibinfo {author} {\bibfnamefont {A.~O.}\ \bibnamefont {Macchiavelli}}, \bibinfo
  {author} {\bibfnamefont {W.}~\bibnamefont {Reviol}}, \bibinfo {author} {\bibfnamefont {D.~G.}\ \bibnamefont {Sarantites}}, \bibinfo {author} {\bibfnamefont {D.}~\bibnamefont {Seweryniak}}, \bibinfo {author} {\bibfnamefont {C.~E.}\ \bibnamefont {Svensson}},\ and\ \bibinfo {author} {\bibfnamefont {S.~J.}\ \bibnamefont {Williams}},\ }\href {https://doi.org/10.1140/epja/i2005-10259-y} {\bibfield  {journal} {\bibinfo  {journal} {Eur. Phys. J. A}\ }\textbf {\bibinfo {volume} {27}},\ \bibinfo {pages} {157} (\bibinfo {year} {2006})}\BibitemShut {NoStop}%
\bibitem [{\citenamefont {Johansson}\ \emph {et~al.}(2008)\citenamefont {Johansson}, \citenamefont {Rudolph}, \citenamefont {Andersson}, \citenamefont {Torres}, \citenamefont {Ragnarsson}, \citenamefont {Andreoiu}, \citenamefont {Baktash}, \citenamefont {Carpenter}, \citenamefont {Charity}, \citenamefont {Chiara}, \citenamefont {Ekman}, \citenamefont {Fahlander}, \citenamefont {Hoel}, \citenamefont {Pechenaya}, \citenamefont {Reviol}, \citenamefont {du~Rietz}, \citenamefont {Sarantites}, \citenamefont {Seweryniak}, \citenamefont {Sobotka}, \citenamefont {Yu},\ and\ \citenamefont {Zhu}}]{RN620}%
  \BibitemOpen
  \bibfield  {author} {\bibinfo {author} {\bibfnamefont {E.~K.}\ \bibnamefont {Johansson}}, \bibinfo {author} {\bibfnamefont {D.}~\bibnamefont {Rudolph}}, \bibinfo {author} {\bibfnamefont {L.-L.}\ \bibnamefont {Andersson}}, \bibinfo {author} {\bibfnamefont {D.~A.}\ \bibnamefont {Torres}}, \bibinfo {author} {\bibfnamefont {I.}~\bibnamefont {Ragnarsson}}, \bibinfo {author} {\bibfnamefont {C.}~\bibnamefont {Andreoiu}}, \bibinfo {author} {\bibfnamefont {C.}~\bibnamefont {Baktash}}, \bibinfo {author} {\bibfnamefont {M.~P.}\ \bibnamefont {Carpenter}}, \bibinfo {author} {\bibfnamefont {R.~J.}\ \bibnamefont {Charity}}, \bibinfo {author} {\bibfnamefont {C.~J.}\ \bibnamefont {Chiara}}, \bibinfo {author} {\bibfnamefont {J.}~\bibnamefont {Ekman}}, \bibinfo {author} {\bibfnamefont {C.}~\bibnamefont {Fahlander}}, \bibinfo {author} {\bibfnamefont {C.}~\bibnamefont {Hoel}}, \bibinfo {author} {\bibfnamefont {O.~L.}\ \bibnamefont {Pechenaya}}, \bibinfo {author} {\bibfnamefont {W.}~\bibnamefont {Reviol}}, \bibinfo {author}
  {\bibfnamefont {R.}~\bibnamefont {du~Rietz}}, \bibinfo {author} {\bibfnamefont {D.~G.}\ \bibnamefont {Sarantites}}, \bibinfo {author} {\bibfnamefont {D.}~\bibnamefont {Seweryniak}}, \bibinfo {author} {\bibfnamefont {L.~G.}\ \bibnamefont {Sobotka}}, \bibinfo {author} {\bibfnamefont {C.~H.}\ \bibnamefont {Yu}},\ and\ \bibinfo {author} {\bibfnamefont {S.}~\bibnamefont {Zhu}},\ }\href {https://doi.org/10.1103/PhysRevC.77.064316} {\bibfield  {journal} {\bibinfo  {journal} {Phys. Rev. C}\ }\textbf {\bibinfo {volume} {77}},\ \bibinfo {pages} {064316} (\bibinfo {year} {2008})}\BibitemShut {NoStop}%
\bibitem [{\citenamefont {Mizusaki}\ \emph {et~al.}(1999)\citenamefont {Mizusaki}, \citenamefont {Otsuka}, \citenamefont {Utsuno}, \citenamefont {Honma},\ and\ \citenamefont {Sebe}}]{RN633}%
  \BibitemOpen
  \bibfield  {author} {\bibinfo {author} {\bibfnamefont {T.}~\bibnamefont {Mizusaki}}, \bibinfo {author} {\bibfnamefont {T.}~\bibnamefont {Otsuka}}, \bibinfo {author} {\bibfnamefont {Y.}~\bibnamefont {Utsuno}}, \bibinfo {author} {\bibfnamefont {M.}~\bibnamefont {Honma}},\ and\ \bibinfo {author} {\bibfnamefont {T.}~\bibnamefont {Sebe}},\ }\href {https://doi.org/10.1103/PhysRevC.59.R1846} {\bibfield  {journal} {\bibinfo  {journal} {Phys. Rev. C}\ }\textbf {\bibinfo {volume} {59}},\ \bibinfo {pages} {R1846} (\bibinfo {year} {1999})}\BibitemShut {NoStop}%
\bibitem [{\citenamefont {Honma}\ \emph {et~al.}(2004)\citenamefont {Honma}, \citenamefont {Otsuka}, \citenamefont {Brown},\ and\ \citenamefont {Mizusaki}}]{honma02prc}%
  \BibitemOpen
  \bibfield  {author} {\bibinfo {author} {\bibfnamefont {M.}~\bibnamefont {Honma}}, \bibinfo {author} {\bibfnamefont {T.}~\bibnamefont {Otsuka}}, \bibinfo {author} {\bibfnamefont {B.~A.}\ \bibnamefont {Brown}},\ and\ \bibinfo {author} {\bibfnamefont {T.}~\bibnamefont {Mizusaki}},\ }\href {https://doi.org/10.1103/PhysRevC.69.034335} {\bibfield  {journal} {\bibinfo  {journal} {Phys. Rev. C}\ }\textbf {\bibinfo {volume} {69}},\ \bibinfo {pages} {034335} (\bibinfo {year} {2004})}\BibitemShut {NoStop}%
\bibitem [{\citenamefont {Petrovici}\ \emph {et~al.}(2001)\citenamefont {Petrovici}, \citenamefont {Schmid},\ and\ \citenamefont {Faessler}}]{RN629}%
  \BibitemOpen
  \bibfield  {author} {\bibinfo {author} {\bibfnamefont {A.}~\bibnamefont {Petrovici}}, \bibinfo {author} {\bibfnamefont {K.~W.}\ \bibnamefont {Schmid}},\ and\ \bibinfo {author} {\bibfnamefont {A.}~\bibnamefont {Faessler}},\ }\href {https://doi.org/Doi 10.1016/S0375-9474(00)00691-6} {\bibfield  {journal} {\bibinfo  {journal} {Nucl. Phys. A}\ }\textbf {\bibinfo {volume} {689}},\ \bibinfo {pages} {707} (\bibinfo {year} {2001})}\BibitemShut {NoStop}%
\bibitem [{\citenamefont {Mizusaki}\ \emph {et~al.}(2002)\citenamefont {Mizusaki}, \citenamefont {Otsuka}, \citenamefont {Honma},\ and\ \citenamefont {Brown}}]{RN630}%
  \BibitemOpen
  \bibfield  {author} {\bibinfo {author} {\bibfnamefont {T.}~\bibnamefont {Mizusaki}}, \bibinfo {author} {\bibfnamefont {T.}~\bibnamefont {Otsuka}}, \bibinfo {author} {\bibfnamefont {M.}~\bibnamefont {Honma}},\ and\ \bibinfo {author} {\bibfnamefont {B.~A.}\ \bibnamefont {Brown}},\ }\href {https://doi.org/https://doi.org/10.1016/S0375-9474(02)00779-0} {\bibfield  {journal} {\bibinfo  {journal} {Nucl. Phys. A}\ }\textbf {\bibinfo {volume} {704}},\ \bibinfo {pages} {190c} (\bibinfo {year} {2002})}\BibitemShut {NoStop}%
\bibitem [{\citenamefont {Nowacki}(2002)}]{RN632}%
  \BibitemOpen
  \bibfield  {author} {\bibinfo {author} {\bibfnamefont {F.}~\bibnamefont {Nowacki}},\ }\href {https://doi.org/https://doi.org/10.1016/S0375-9474(02)00782-0} {\bibfield  {journal} {\bibinfo  {journal} {Nucl. Phys. A}\ }\textbf {\bibinfo {volume} {704}},\ \bibinfo {pages} {223c} (\bibinfo {year} {2002})}\BibitemShut {NoStop}%
\bibitem [{\citenamefont {Horoi}\ \emph {et~al.}(2003)\citenamefont {Horoi}, \citenamefont {Brown},\ and\ \citenamefont {Zelevinsky}}]{RN652}%
  \BibitemOpen
  \bibfield  {author} {\bibinfo {author} {\bibfnamefont {M.}~\bibnamefont {Horoi}}, \bibinfo {author} {\bibfnamefont {B.~A.}\ \bibnamefont {Brown}},\ and\ \bibinfo {author} {\bibfnamefont {V.}~\bibnamefont {Zelevinsky}},\ }\href {https://doi.org/10.1103/PhysRevC.67.034303} {\bibfield  {journal} {\bibinfo  {journal} {Phys. Rev. C}\ }\textbf {\bibinfo {volume} {67}},\ \bibinfo {pages} {034303} (\bibinfo {year} {2003})}\BibitemShut {NoStop}%
\bibitem [{\citenamefont {Jiang}\ \emph {et~al.}(2003)\citenamefont {Jiang}, \citenamefont {Wang},\ and\ \citenamefont {Zhu}}]{RN654}%
  \BibitemOpen
  \bibfield  {author} {\bibinfo {author} {\bibfnamefont {W.~Z.}\ \bibnamefont {Jiang}}, \bibinfo {author} {\bibfnamefont {T.~T.}\ \bibnamefont {Wang}},\ and\ \bibinfo {author} {\bibfnamefont {Z.~Y.}\ \bibnamefont {Zhu}},\ }\href {https://doi.org/10.1103/PhysRevC.68.047301} {\bibfield  {journal} {\bibinfo  {journal} {Phys. Rev. C}\ }\textbf {\bibinfo {volume} {68}},\ \bibinfo {pages} {047301} (\bibinfo {year} {2003})}\BibitemShut {NoStop}%
\bibitem [{\citenamefont {Dong}\ and\ \citenamefont {Guo}(2004)}]{RN644}%
  \BibitemOpen
  \bibfield  {author} {\bibinfo {author} {\bibfnamefont {B.~G.}\ \bibnamefont {Dong}}\ and\ \bibinfo {author} {\bibfnamefont {H.~C.}\ \bibnamefont {Guo}},\ }\href@noop {} {\bibfield  {journal} {\bibinfo  {journal} {Chin. Phys. Lett.}\ }\textbf {\bibinfo {volume} {21}},\ \bibinfo {pages} {2144} (\bibinfo {year} {2004})}\BibitemShut {NoStop}%
\bibitem [{\citenamefont {Horoi}\ \emph {et~al.}(2006)\citenamefont {Horoi}, \citenamefont {Brown}, \citenamefont {Otsuka}, \citenamefont {Honma},\ and\ \citenamefont {Mizusaki}}]{RN636}%
  \BibitemOpen
  \bibfield  {author} {\bibinfo {author} {\bibfnamefont {M.}~\bibnamefont {Horoi}}, \bibinfo {author} {\bibfnamefont {B.~A.}\ \bibnamefont {Brown}}, \bibinfo {author} {\bibfnamefont {T.}~\bibnamefont {Otsuka}}, \bibinfo {author} {\bibfnamefont {M.}~\bibnamefont {Honma}},\ and\ \bibinfo {author} {\bibfnamefont {T.}~\bibnamefont {Mizusaki}},\ }\href {https://doi.org/10.1103/PhysRevC.73.061305} {\bibfield  {journal} {\bibinfo  {journal} {Phys. Rev. C}\ }\textbf {\bibinfo {volume} {73}},\ \bibinfo {pages} {061305} (\bibinfo {year} {2006})}\BibitemShut {NoStop}%
\bibitem [{\citenamefont {Nazarewicz}\ \emph {et~al.}(1984)\citenamefont {Nazarewicz}, \citenamefont {Olanders}, \citenamefont {Ragnarsson}, \citenamefont {Dudek}, \citenamefont {Leander}, \citenamefont {Möller},\ and\ \citenamefont {Ruchowsa}}]{NPA1984}%
  \BibitemOpen
  \bibfield  {author} {\bibinfo {author} {\bibfnamefont {W.}~\bibnamefont {Nazarewicz}}, \bibinfo {author} {\bibfnamefont {P.}~\bibnamefont {Olanders}}, \bibinfo {author} {\bibfnamefont {I.}~\bibnamefont {Ragnarsson}}, \bibinfo {author} {\bibfnamefont {J.}~\bibnamefont {Dudek}}, \bibinfo {author} {\bibfnamefont {G.}~\bibnamefont {Leander}}, \bibinfo {author} {\bibfnamefont {P.}~\bibnamefont {Möller}},\ and\ \bibinfo {author} {\bibfnamefont {E.}~\bibnamefont {Ruchowsa}},\ }\href {https://doi.org/https://doi.org/10.1016/0375-9474(84)90208-2} {\bibfield  {journal} {\bibinfo  {journal} {Nucl. Phys. A}\ }\textbf {\bibinfo {volume} {429}},\ \bibinfo {pages} {269} (\bibinfo {year} {1984})}\BibitemShut {NoStop}%
\bibitem [{\citenamefont {Möller}\ \emph {et~al.}(2008)\citenamefont {Möller}, \citenamefont {Bengtsson}, \citenamefont {Carlsson}, \citenamefont {Olivius}, \citenamefont {Ichikawa}, \citenamefont {Sagawa},\ and\ \citenamefont {Iwamoto}}]{MOLLER2008758}%
  \BibitemOpen
  \bibfield  {author} {\bibinfo {author} {\bibfnamefont {P.}~\bibnamefont {Möller}}, \bibinfo {author} {\bibfnamefont {R.}~\bibnamefont {Bengtsson}}, \bibinfo {author} {\bibfnamefont {B.}~\bibnamefont {Carlsson}}, \bibinfo {author} {\bibfnamefont {P.}~\bibnamefont {Olivius}}, \bibinfo {author} {\bibfnamefont {T.}~\bibnamefont {Ichikawa}}, \bibinfo {author} {\bibfnamefont {H.}~\bibnamefont {Sagawa}},\ and\ \bibinfo {author} {\bibfnamefont {A.}~\bibnamefont {Iwamoto}},\ }\href {https://doi.org/https://doi.org/10.1016/j.adt.2008.05.002} {\bibfield  {journal} {\bibinfo  {journal} {At. Data Nucl. Data Tables}\ }\textbf {\bibinfo {volume} {94}},\ \bibinfo {pages} {758} (\bibinfo {year} {2008})}\BibitemShut {NoStop}%
\bibitem [{\citenamefont {Jachimowicz}\ \emph {et~al.}(2017)\citenamefont {Jachimowicz}, \citenamefont {Kowal},\ and\ \citenamefont {Skalski}}]{Kowalprc17}%
  \BibitemOpen
  \bibfield  {author} {\bibinfo {author} {\bibfnamefont {P.}~\bibnamefont {Jachimowicz}}, \bibinfo {author} {\bibfnamefont {M.}~\bibnamefont {Kowal}},\ and\ \bibinfo {author} {\bibfnamefont {J.}~\bibnamefont {Skalski}},\ }\href {https://doi.org/10.1103/PhysRevC.95.034329} {\bibfield  {journal} {\bibinfo  {journal} {Phys. Rev. C}\ }\textbf {\bibinfo {volume} {95}},\ \bibinfo {pages} {034329} (\bibinfo {year} {2017})}\BibitemShut {NoStop}%
\bibitem [{\citenamefont {Nerlo-Pomorska}\ \emph {et~al.}(2017)\citenamefont {Nerlo-Pomorska}, \citenamefont {Pomorski}, \citenamefont {Bartel},\ and\ \citenamefont {Schmitt}}]{RN942}%
  \BibitemOpen
  \bibfield  {author} {\bibinfo {author} {\bibfnamefont {B.}~\bibnamefont {Nerlo-Pomorska}}, \bibinfo {author} {\bibfnamefont {K.}~\bibnamefont {Pomorski}}, \bibinfo {author} {\bibfnamefont {J.}~\bibnamefont {Bartel}},\ and\ \bibinfo {author} {\bibfnamefont {C.}~\bibnamefont {Schmitt}},\ }\href {https://doi.org/10.1140/epja/i2017-12259-8} {\bibfield  {journal} {\bibinfo  {journal} {Eur. Phys. J}\ }\textbf {\bibinfo {volume} {53}},\ \bibinfo {pages} {67} (\bibinfo {year} {2017})}\BibitemShut {NoStop}%
\bibitem [{\citenamefont {Dobaczewski}\ and\ \citenamefont {Engel}(2005)}]{Dobaczewskiprl05}%
  \BibitemOpen
  \bibfield  {author} {\bibinfo {author} {\bibfnamefont {J.}~\bibnamefont {Dobaczewski}}\ and\ \bibinfo {author} {\bibfnamefont {J.}~\bibnamefont {Engel}},\ }\href {https://doi.org/10.1103/PhysRevLett.94.232502} {\bibfield  {journal} {\bibinfo  {journal} {Phys. Rev. Lett.}\ }\textbf {\bibinfo {volume} {94}},\ \bibinfo {pages} {232502} (\bibinfo {year} {2005})}\BibitemShut {NoStop}%
\bibitem [{\citenamefont {Robledo}\ and\ \citenamefont {Bertsch}(2011)}]{Robledoprc11}%
  \BibitemOpen
  \bibfield  {author} {\bibinfo {author} {\bibfnamefont {L.~M.}\ \bibnamefont {Robledo}}\ and\ \bibinfo {author} {\bibfnamefont {G.~F.}\ \bibnamefont {Bertsch}},\ }\href {https://doi.org/10.1103/PhysRevC.84.054302} {\bibfield  {journal} {\bibinfo  {journal} {Phys. Rev. C}\ }\textbf {\bibinfo {volume} {84}},\ \bibinfo {pages} {054302} (\bibinfo {year} {2011})}\BibitemShut {NoStop}%
\bibitem [{\citenamefont {Robledo}\ and\ \citenamefont {Rodríguez-Guzmán}(2012)}]{Robledo_2012}%
  \BibitemOpen
  \bibfield  {author} {\bibinfo {author} {\bibfnamefont {L.~M.}\ \bibnamefont {Robledo}}\ and\ \bibinfo {author} {\bibfnamefont {R.~R.}\ \bibnamefont {Rodríguez-Guzmán}},\ }\href {https://doi.org/10.1088/0954-3899/39/10/105103} {\bibfield  {journal} {\bibinfo  {journal} {J. Phys. G: Nucl. Part. Phys.}\ }\textbf {\bibinfo {volume} {39}},\ \bibinfo {pages} {105103} (\bibinfo {year} {2012})}\BibitemShut {NoStop}%
\bibitem [{\citenamefont {Li}\ \emph {et~al.}(2013)\citenamefont {Li}, \citenamefont {Song}, \citenamefont {Yao}, \citenamefont {Vretenar},\ and\ \citenamefont {Meng}}]{LI2013866}%
  \BibitemOpen
  \bibfield  {author} {\bibinfo {author} {\bibfnamefont {Z.}~\bibnamefont {Li}}, \bibinfo {author} {\bibfnamefont {B.}~\bibnamefont {Song}}, \bibinfo {author} {\bibfnamefont {J.}~\bibnamefont {Yao}}, \bibinfo {author} {\bibfnamefont {D.}~\bibnamefont {Vretenar}},\ and\ \bibinfo {author} {\bibfnamefont {J.}~\bibnamefont {Meng}},\ }\href {https://doi.org/https://doi.org/10.1016/j.physletb.2013.09.035} {\bibfield  {journal} {\bibinfo  {journal} {Phys. Lett. B}\ }\textbf {\bibinfo {volume} {726}},\ \bibinfo {pages} {866} (\bibinfo {year} {2013})}\BibitemShut {NoStop}%
\bibitem [{\citenamefont {Lu}\ \emph {et~al.}(2014{\natexlab{a}})\citenamefont {Lu}, \citenamefont {Zhao}, \citenamefont {Zhao},\ and\ \citenamefont {Zhou}}]{luprc14}%
  \BibitemOpen
  \bibfield  {author} {\bibinfo {author} {\bibfnamefont {B.-N.}\ \bibnamefont {Lu}}, \bibinfo {author} {\bibfnamefont {J.}~\bibnamefont {Zhao}}, \bibinfo {author} {\bibfnamefont {E.-G.}\ \bibnamefont {Zhao}},\ and\ \bibinfo {author} {\bibfnamefont {S.-G.}\ \bibnamefont {Zhou}},\ }\href {https://doi.org/10.1103/PhysRevC.89.014323} {\bibfield  {journal} {\bibinfo  {journal} {Phys. Rev. C}\ }\textbf {\bibinfo {volume} {89}},\ \bibinfo {pages} {014323} (\bibinfo {year} {2014}{\natexlab{a}})}\BibitemShut {NoStop}%
\bibitem [{\citenamefont {Zhou}(2016)}]{Zhou_2016}%
  \BibitemOpen
  \bibfield  {author} {\bibinfo {author} {\bibfnamefont {S.-G.}\ \bibnamefont {Zhou}},\ }\href {https://doi.org/10.1088/0031-8949/91/6/063008} {\bibfield  {journal} {\bibinfo  {journal} {Phys. Scr.}\ }\textbf {\bibinfo {volume} {91}},\ \bibinfo {pages} {063008} (\bibinfo {year} {2016})}\BibitemShut {NoStop}%
\bibitem [{\citenamefont {Wang}\ and\ \citenamefont {Lu}(2022)}]{Wangk2022}%
  \BibitemOpen
  \bibfield  {author} {\bibinfo {author} {\bibfnamefont {K.}~\bibnamefont {Wang}}\ and\ \bibinfo {author} {\bibfnamefont {B.-N.}\ \bibnamefont {Lu}},\ }\href {https://doi.org/10.1088/1572-9494/ac3999} {\bibfield  {journal} {\bibinfo  {journal} {Commun. Theor. Phys.}\ }\textbf {\bibinfo {volume} {74}},\ \bibinfo {pages} {015303} (\bibinfo {year} {2022})}\BibitemShut {NoStop}%
\bibitem [{\citenamefont {Chen}\ \emph {et~al.}(2015)\citenamefont {Chen}, \citenamefont {Gao}, \citenamefont {Chen},\ and\ \citenamefont {Tu}}]{Chenprc15}%
  \BibitemOpen
  \bibfield  {author} {\bibinfo {author} {\bibfnamefont {Y.-J.}\ \bibnamefont {Chen}}, \bibinfo {author} {\bibfnamefont {Z.-C.}\ \bibnamefont {Gao}}, \bibinfo {author} {\bibfnamefont {Y.-S.}\ \bibnamefont {Chen}},\ and\ \bibinfo {author} {\bibfnamefont {Y.}~\bibnamefont {Tu}},\ }\href {https://doi.org/10.1103/PhysRevC.91.014317} {\bibfield  {journal} {\bibinfo  {journal} {Phys. Rev. C}\ }\textbf {\bibinfo {volume} {91}},\ \bibinfo {pages} {014317} (\bibinfo {year} {2015})}\BibitemShut {NoStop}%
\bibitem [{\citenamefont {Agbemava}\ \emph {et~al.}(2016)\citenamefont {Agbemava}, \citenamefont {Afanasjev},\ and\ \citenamefont {Ring}}]{Agbemavaprc16}%
  \BibitemOpen
  \bibfield  {author} {\bibinfo {author} {\bibfnamefont {S.~E.}\ \bibnamefont {Agbemava}}, \bibinfo {author} {\bibfnamefont {A.~V.}\ \bibnamefont {Afanasjev}},\ and\ \bibinfo {author} {\bibfnamefont {P.}~\bibnamefont {Ring}},\ }\href {https://doi.org/10.1103/PhysRevC.93.044304} {\bibfield  {journal} {\bibinfo  {journal} {Phys. Rev. C}\ }\textbf {\bibinfo {volume} {93}},\ \bibinfo {pages} {044304} (\bibinfo {year} {2016})}\BibitemShut {NoStop}%
\bibitem [{\citenamefont {Nomura}\ \emph {et~al.}(2018)\citenamefont {Nomura}, \citenamefont {Nik\ifmmode \check{s}\else \v{s}\fi{}i\ifmmode~\acute{c}\else \'{c}\fi{}},\ and\ \citenamefont {Vretenar}}]{Nomuraprc18}%
  \BibitemOpen
  \bibfield  {author} {\bibinfo {author} {\bibfnamefont {K.}~\bibnamefont {Nomura}}, \bibinfo {author} {\bibfnamefont {T.}~\bibnamefont {Nik\ifmmode \check{s}\else \v{s}\fi{}i\ifmmode~\acute{c}\else \'{c}\fi{}}},\ and\ \bibinfo {author} {\bibfnamefont {D.}~\bibnamefont {Vretenar}},\ }\href {https://doi.org/10.1103/PhysRevC.97.024317} {\bibfield  {journal} {\bibinfo  {journal} {Phys. Rev. C}\ }\textbf {\bibinfo {volume} {97}},\ \bibinfo {pages} {024317} (\bibinfo {year} {2018})}\BibitemShut {NoStop}%
\bibitem [{\citenamefont {Ganev}(2019)}]{Ganev19}%
  \BibitemOpen
  \bibfield  {author} {\bibinfo {author} {\bibfnamefont {H.~G.}\ \bibnamefont {Ganev}},\ }\href {https://doi.org/10.1103/PhysRevC.99.054305} {\bibfield  {journal} {\bibinfo  {journal} {Phys. Rev. C}\ }\textbf {\bibinfo {volume} {99}},\ \bibinfo {pages} {054305} (\bibinfo {year} {2019})}\BibitemShut {NoStop}%
\bibitem [{\citenamefont {Sun}\ \emph {et~al.}(2019)\citenamefont {Sun}, \citenamefont {Quan}, \citenamefont {Li}, \citenamefont {Zhao}, \citenamefont {Nik\ifmmode \check{s}\else \v{s}\fi{}i\ifmmode~\acute{c}\else \'{c}\fi{}},\ and\ \citenamefont {Vretenar}}]{Lizp19}%
  \BibitemOpen
  \bibfield  {author} {\bibinfo {author} {\bibfnamefont {W.}~\bibnamefont {Sun}}, \bibinfo {author} {\bibfnamefont {S.}~\bibnamefont {Quan}}, \bibinfo {author} {\bibfnamefont {Z.~P.}\ \bibnamefont {Li}}, \bibinfo {author} {\bibfnamefont {J.}~\bibnamefont {Zhao}}, \bibinfo {author} {\bibfnamefont {T.}~\bibnamefont {Nik\ifmmode \check{s}\else \v{s}\fi{}i\ifmmode~\acute{c}\else \'{c}\fi{}}},\ and\ \bibinfo {author} {\bibfnamefont {D.}~\bibnamefont {Vretenar}},\ }\href {https://doi.org/10.1103/PhysRevC.100.044319} {\bibfield  {journal} {\bibinfo  {journal} {Phys. Rev. C}\ }\textbf {\bibinfo {volume} {100}},\ \bibinfo {pages} {044319} (\bibinfo {year} {2019})}\BibitemShut {NoStop}%
\bibitem [{\citenamefont {Cao}\ \emph {et~al.}(2020)\citenamefont {Cao}, \citenamefont {Agbemava}, \citenamefont {Afanasjev}, \citenamefont {Nazarewicz},\ and\ \citenamefont {Olsen}}]{cyc20}%
  \BibitemOpen
  \bibfield  {author} {\bibinfo {author} {\bibfnamefont {Y.}~\bibnamefont {Cao}}, \bibinfo {author} {\bibfnamefont {S.~E.}\ \bibnamefont {Agbemava}}, \bibinfo {author} {\bibfnamefont {A.~V.}\ \bibnamefont {Afanasjev}}, \bibinfo {author} {\bibfnamefont {W.}~\bibnamefont {Nazarewicz}},\ and\ \bibinfo {author} {\bibfnamefont {E.}~\bibnamefont {Olsen}},\ }\href {https://doi.org/10.1103/PhysRevC.102.024311} {\bibfield  {journal} {\bibinfo  {journal} {Phys. Rev. C}\ }\textbf {\bibinfo {volume} {102}},\ \bibinfo {pages} {024311} (\bibinfo {year} {2020})}\BibitemShut {NoStop}%
\bibitem [{\citenamefont {He}\ and\ \citenamefont {Li}(2020)}]{hext20}%
  \BibitemOpen
  \bibfield  {author} {\bibinfo {author} {\bibfnamefont {X.-T.}\ \bibnamefont {He}}\ and\ \bibinfo {author} {\bibfnamefont {Y.-C.}\ \bibnamefont {Li}},\ }\href {https://doi.org/10.1103/PhysRevC.102.064328} {\bibfield  {journal} {\bibinfo  {journal} {Phys. Rev. C}\ }\textbf {\bibinfo {volume} {102}},\ \bibinfo {pages} {064328} (\bibinfo {year} {2020})}\BibitemShut {NoStop}%
\bibitem [{\citenamefont {Nomura}\ \emph {et~al.}(2021)\citenamefont {Nomura}, \citenamefont {Rodr\'{\i}guez-Guzm\'an}, \citenamefont {Robledo}, \citenamefont {Garc\'{\i}a-Ramos},\ and\ \citenamefont {Hern\'andez}}]{Nomura21}%
  \BibitemOpen
  \bibfield  {author} {\bibinfo {author} {\bibfnamefont {K.}~\bibnamefont {Nomura}}, \bibinfo {author} {\bibfnamefont {R.}~\bibnamefont {Rodr\'{\i}guez-Guzm\'an}}, \bibinfo {author} {\bibfnamefont {L.~M.}\ \bibnamefont {Robledo}}, \bibinfo {author} {\bibfnamefont {J.~E.}\ \bibnamefont {Garc\'{\i}a-Ramos}},\ and\ \bibinfo {author} {\bibfnamefont {N.~C.}\ \bibnamefont {Hern\'andez}},\ }\href {https://doi.org/10.1103/PhysRevC.104.044324} {\bibfield  {journal} {\bibinfo  {journal} {Phys. Rev. C}\ }\textbf {\bibinfo {volume} {104}},\ \bibinfo {pages} {044324} (\bibinfo {year} {2021})}\BibitemShut {NoStop}%
\bibitem [{\citenamefont {Nomura}(2022)}]{Nomuraprc22}%
  \BibitemOpen
  \bibfield  {author} {\bibinfo {author} {\bibfnamefont {K.}~\bibnamefont {Nomura}},\ }\href {https://doi.org/10.1103/PhysRevC.105.054318} {\bibfield  {journal} {\bibinfo  {journal} {Phys. Rev. C}\ }\textbf {\bibinfo {volume} {105}},\ \bibinfo {pages} {054318} (\bibinfo {year} {2022})}\BibitemShut {NoStop}%
\bibitem [{\citenamefont {Qiu}\ \emph {et~al.}(2022)\citenamefont {Qiu}, \citenamefont {Wang},\ and\ \citenamefont {Guo}}]{Guojyprc22}%
  \BibitemOpen
  \bibfield  {author} {\bibinfo {author} {\bibfnamefont {Y.-T.}\ \bibnamefont {Qiu}}, \bibinfo {author} {\bibfnamefont {X.-W.}\ \bibnamefont {Wang}},\ and\ \bibinfo {author} {\bibfnamefont {J.-Y.}\ \bibnamefont {Guo}},\ }\href {https://doi.org/10.1103/PhysRevC.106.034301} {\bibfield  {journal} {\bibinfo  {journal} {Phys. Rev. C}\ }\textbf {\bibinfo {volume} {106}},\ \bibinfo {pages} {034301} (\bibinfo {year} {2022})}\BibitemShut {NoStop}%
\bibitem [{\citenamefont {Rong}\ \emph {et~al.}(2023)\citenamefont {Rong}, \citenamefont {Wu}, \citenamefont {Lu},\ and\ \citenamefont {Yao}}]{RONG2023}%
  \BibitemOpen
  \bibfield  {author} {\bibinfo {author} {\bibfnamefont {Y.-T.}\ \bibnamefont {Rong}}, \bibinfo {author} {\bibfnamefont {X.-Y.}\ \bibnamefont {Wu}}, \bibinfo {author} {\bibfnamefont {B.-N.}\ \bibnamefont {Lu}},\ and\ \bibinfo {author} {\bibfnamefont {J.-M.}\ \bibnamefont {Yao}},\ }\href {https://doi.org/https://doi.org/10.1016/j.physletb.2023.137896} {\bibfield  {journal} {\bibinfo  {journal} {Phys. Lett. B}\ }\textbf {\bibinfo {volume} {840}},\ \bibinfo {pages} {137896} (\bibinfo {year} {2023})}\BibitemShut {NoStop}%
\bibitem [{\citenamefont {Minh~Loc}\ \emph {et~al.}(2023)\citenamefont {Minh~Loc}, \citenamefont {Le~Anh}, \citenamefont {Papakonstantinou},\ and\ \citenamefont {Auerbach}}]{Minhprc24}%
  \BibitemOpen
  \bibfield  {author} {\bibinfo {author} {\bibfnamefont {B.}~\bibnamefont {Minh~Loc}}, \bibinfo {author} {\bibfnamefont {N.}~\bibnamefont {Le~Anh}}, \bibinfo {author} {\bibfnamefont {P.}~\bibnamefont {Papakonstantinou}},\ and\ \bibinfo {author} {\bibfnamefont {N.}~\bibnamefont {Auerbach}},\ }\href {https://doi.org/10.1103/PhysRevC.108.024303} {\bibfield  {journal} {\bibinfo  {journal} {Phys. Rev. C}\ }\textbf {\bibinfo {volume} {108}},\ \bibinfo {pages} {024303} (\bibinfo {year} {2023})}\BibitemShut {NoStop}%
\bibitem [{\citenamefont {Nomura}(2024)}]{Nomuraprc24}%
  \BibitemOpen
  \bibfield  {author} {\bibinfo {author} {\bibfnamefont {K.}~\bibnamefont {Nomura}},\ }\href {https://doi.org/10.1103/PhysRevC.110.064306} {\bibfield  {journal} {\bibinfo  {journal} {Phys. Rev. C}\ }\textbf {\bibinfo {volume} {110}},\ \bibinfo {pages} {064306} (\bibinfo {year} {2024})}\BibitemShut {NoStop}%
\bibitem [{\citenamefont {Li}\ and\ \citenamefont {Wang}(2024)}]{wang24}%
  \BibitemOpen
  \bibfield  {author} {\bibinfo {author} {\bibfnamefont {Z.-K.}\ \bibnamefont {Li}}\ and\ \bibinfo {author} {\bibfnamefont {Y.-Y.}\ \bibnamefont {Wang}},\ }\href {https://doi.org/10.1007/s41365-024-01532-z} {\bibfield  {journal} {\bibinfo  {journal} {Nucl. Sci. Tech.}\ }\textbf {\bibinfo {volume} {35}},\ \bibinfo {pages} {139} (\bibinfo {year} {2024})}\BibitemShut {NoStop}%
\bibitem [{\citenamefont {Yin}\ \emph {et~al.}(2024)\citenamefont {Yin}, \citenamefont {Ma},\ and\ \citenamefont {Zhao}}]{zhaoymprc24}%
  \BibitemOpen
  \bibfield  {author} {\bibinfo {author} {\bibfnamefont {X.}~\bibnamefont {Yin}}, \bibinfo {author} {\bibfnamefont {C.}~\bibnamefont {Ma}},\ and\ \bibinfo {author} {\bibfnamefont {Y.~M.}\ \bibnamefont {Zhao}},\ }\href {https://doi.org/10.1103/PhysRevC.109.024322} {\bibfield  {journal} {\bibinfo  {journal} {Phys. Rev. C}\ }\textbf {\bibinfo {volume} {109}},\ \bibinfo {pages} {024322} (\bibinfo {year} {2024})}\BibitemShut {NoStop}%
\bibitem [{\citenamefont {Xu}\ \emph {et~al.}(2024)\citenamefont {Xu}, \citenamefont {Li}, \citenamefont {Ring},\ and\ \citenamefont {Zhao}}]{XU2024138893}%
  \BibitemOpen
  \bibfield  {author} {\bibinfo {author} {\bibfnamefont {F.}~\bibnamefont {Xu}}, \bibinfo {author} {\bibfnamefont {B.}~\bibnamefont {Li}}, \bibinfo {author} {\bibfnamefont {P.}~\bibnamefont {Ring}},\ and\ \bibinfo {author} {\bibfnamefont {P.}~\bibnamefont {Zhao}},\ }\href {https://doi.org/https://doi.org/10.1016/j.physletb.2024.138893} {\bibfield  {journal} {\bibinfo  {journal} {Phys. Lett. B}\ }\textbf {\bibinfo {volume} {856}},\ \bibinfo {pages} {138893} (\bibinfo {year} {2024})}\BibitemShut {NoStop}%
\bibitem [{\citenamefont {Shneidman}\ \emph {et~al.}(2003)\citenamefont {Shneidman}, \citenamefont {Adamian}, \citenamefont {Antonenko}, \citenamefont {Jolos},\ and\ \citenamefont {Scheid}}]{Shneidmanprc03}%
  \BibitemOpen
  \bibfield  {author} {\bibinfo {author} {\bibfnamefont {T.~M.}\ \bibnamefont {Shneidman}}, \bibinfo {author} {\bibfnamefont {G.~G.}\ \bibnamefont {Adamian}}, \bibinfo {author} {\bibfnamefont {N.~V.}\ \bibnamefont {Antonenko}}, \bibinfo {author} {\bibfnamefont {R.~V.}\ \bibnamefont {Jolos}},\ and\ \bibinfo {author} {\bibfnamefont {W.}~\bibnamefont {Scheid}},\ }\href {https://doi.org/10.1103/PhysRevC.67.014313} {\bibfield  {journal} {\bibinfo  {journal} {Phys. Rev. C}\ }\textbf {\bibinfo {volume} {67}},\ \bibinfo {pages} {014313} (\bibinfo {year} {2003})}\BibitemShut {NoStop}%
\bibitem [{\citenamefont {Buck}\ \emph {et~al.}(2008)\citenamefont {Buck}, \citenamefont {Merchant},\ and\ \citenamefont {Perez}}]{Buck2008}%
  \BibitemOpen
  \bibfield  {author} {\bibinfo {author} {\bibfnamefont {B.}~\bibnamefont {Buck}}, \bibinfo {author} {\bibfnamefont {A.~C.}\ \bibnamefont {Merchant}},\ and\ \bibinfo {author} {\bibfnamefont {S.~M.}\ \bibnamefont {Perez}},\ }\href {https://api.semanticscholar.org/CorpusID:121087145} {\bibfield  {journal} {\bibinfo  {journal} {J. Phys. G}\ }\textbf {\bibinfo {volume} {35}},\ \bibinfo {pages} {085101} (\bibinfo {year} {2008})}\BibitemShut {NoStop}%
\bibitem [{\citenamefont {Zamfir}\ and\ \citenamefont {Kusnezov}(2001)}]{Zamfirprc01}%
  \BibitemOpen
  \bibfield  {author} {\bibinfo {author} {\bibfnamefont {N.~V.}\ \bibnamefont {Zamfir}}\ and\ \bibinfo {author} {\bibfnamefont {D.}~\bibnamefont {Kusnezov}},\ }\href {https://doi.org/10.1103/PhysRevC.63.054306} {\bibfield  {journal} {\bibinfo  {journal} {Phys. Rev. C}\ }\textbf {\bibinfo {volume} {63}},\ \bibinfo {pages} {054306} (\bibinfo {year} {2001})}\BibitemShut {NoStop}%
\bibitem [{\citenamefont {Lu}\ \emph {et~al.}(2018)\citenamefont {Lu}, \citenamefont {Qi},\ and\ \citenamefont {Wang}}]{Lu_2018}%
  \BibitemOpen
  \bibfield  {author} {\bibinfo {author} {\bibfnamefont {X.}~\bibnamefont {Lu}}, \bibinfo {author} {\bibfnamefont {B.}~\bibnamefont {Qi}},\ and\ \bibinfo {author} {\bibfnamefont {S.-Y.}\ \bibnamefont {Wang}},\ }\href {https://doi.org/10.1088/0256-307X/35/10/102101} {\bibfield  {journal} {\bibinfo  {journal} {Chin. Phys. Lett.}\ }\textbf {\bibinfo {volume} {35}},\ \bibinfo {pages} {102101} (\bibinfo {year} {2018})}\BibitemShut {NoStop}%
\bibitem [{\citenamefont {Wang}\ \emph {et~al.}(2019)\citenamefont {Wang}, \citenamefont {Zhang}, \citenamefont {Zhao},\ and\ \citenamefont {Meng}}]{WANG2019454}%
  \BibitemOpen
  \bibfield  {author} {\bibinfo {author} {\bibfnamefont {Y.}~\bibnamefont {Wang}}, \bibinfo {author} {\bibfnamefont {S.}~\bibnamefont {Zhang}}, \bibinfo {author} {\bibfnamefont {P.}~\bibnamefont {Zhao}},\ and\ \bibinfo {author} {\bibfnamefont {J.}~\bibnamefont {Meng}},\ }\href {https://doi.org/https://doi.org/10.1016/j.physletb.2019.04.014} {\bibfield  {journal} {\bibinfo  {journal} {Phys. Lett. B}\ }\textbf {\bibinfo {volume} {792}},\ \bibinfo {pages} {454} (\bibinfo {year} {2019})}\BibitemShut {NoStop}%
\bibitem [{\citenamefont {Nadirbekov}\ \emph {et~al.}(2022)\citenamefont {Nadirbekov}, \citenamefont {Bozarov}, \citenamefont {Kudiratov},\ and\ \citenamefont {Minkov}}]{Minkov22}%
  \BibitemOpen
  \bibfield  {author} {\bibinfo {author} {\bibfnamefont {M.~S.}\ \bibnamefont {Nadirbekov}}, \bibinfo {author} {\bibfnamefont {O.~A.}\ \bibnamefont {Bozarov}}, \bibinfo {author} {\bibfnamefont {S.~N.}\ \bibnamefont {Kudiratov}},\ and\ \bibinfo {author} {\bibfnamefont {N.}~\bibnamefont {Minkov}},\ }\href {https://doi.org/10.1142/S0218301322500781} {\bibfield  {journal} {\bibinfo  {journal} {Int. J. Mod. Phys. E}\ }\textbf {\bibinfo {volume} {31}},\ \bibinfo {pages} {2250078} (\bibinfo {year} {2022})}\BibitemShut {NoStop}%
\bibitem [{\citenamefont {Wang}\ \emph {et~al.}(2022{\natexlab{b}})\citenamefont {Wang}, \citenamefont {Qi}, \citenamefont {Liu}, \citenamefont {Rohilla},\ and\ \citenamefont {Zhang}}]{qibin22}%
  \BibitemOpen
  \bibfield  {author} {\bibinfo {author} {\bibfnamefont {X.~D.}\ \bibnamefont {Wang}}, \bibinfo {author} {\bibfnamefont {B.}~\bibnamefont {Qi}}, \bibinfo {author} {\bibfnamefont {C.}~\bibnamefont {Liu}}, \bibinfo {author} {\bibfnamefont {A.}~\bibnamefont {Rohilla}},\ and\ \bibinfo {author} {\bibfnamefont {Y.}~\bibnamefont {Zhang}},\ }\href {https://doi.org/10.1103/PhysRevC.106.064325} {\bibfield  {journal} {\bibinfo  {journal} {Phys. Rev. C}\ }\textbf {\bibinfo {volume} {106}},\ \bibinfo {pages} {064325} (\bibinfo {year} {2022}{\natexlab{b}})}\BibitemShut {NoStop}%
\bibitem [{\citenamefont {Wang}\ \emph {et~al.}(2022{\natexlab{c}})\citenamefont {Wang}, \citenamefont {Chen},\ and\ \citenamefont {Zhang}}]{wangyyprc22}%
  \BibitemOpen
  \bibfield  {author} {\bibinfo {author} {\bibfnamefont {Y.~Y.}\ \bibnamefont {Wang}}, \bibinfo {author} {\bibfnamefont {Q.~B.}\ \bibnamefont {Chen}},\ and\ \bibinfo {author} {\bibfnamefont {S.~Q.}\ \bibnamefont {Zhang}},\ }\href {https://doi.org/10.1103/PhysRevC.105.044316} {\bibfield  {journal} {\bibinfo  {journal} {Phys. Rev. C}\ }\textbf {\bibinfo {volume} {105}},\ \bibinfo {pages} {044316} (\bibinfo {year} {2022}{\natexlab{c}})}\BibitemShut {NoStop}%
\bibitem [{\citenamefont {Jolos}\ \emph {et~al.}(2012)\citenamefont {Jolos}, \citenamefont {von Brentano},\ and\ \citenamefont {Jolie}}]{Jolosprc12}%
  \BibitemOpen
  \bibfield  {author} {\bibinfo {author} {\bibfnamefont {R.~V.}\ \bibnamefont {Jolos}}, \bibinfo {author} {\bibfnamefont {P.}~\bibnamefont {von Brentano}},\ and\ \bibinfo {author} {\bibfnamefont {J.}~\bibnamefont {Jolie}},\ }\href {https://doi.org/10.1103/PhysRevC.86.024319} {\bibfield  {journal} {\bibinfo  {journal} {Phys. Rev. C}\ }\textbf {\bibinfo {volume} {86}},\ \bibinfo {pages} {024319} (\bibinfo {year} {2012})}\BibitemShut {NoStop}%
\bibitem [{\citenamefont {Seif}\ \emph {et~al.}(2023)\citenamefont {Seif}, \citenamefont {Adel}, \citenamefont {Antonenko},\ and\ \citenamefont {Adamian}}]{Antonenkoprc23}%
  \BibitemOpen
  \bibfield  {author} {\bibinfo {author} {\bibfnamefont {W.~M.}\ \bibnamefont {Seif}}, \bibinfo {author} {\bibfnamefont {A.}~\bibnamefont {Adel}}, \bibinfo {author} {\bibfnamefont {N.~V.}\ \bibnamefont {Antonenko}},\ and\ \bibinfo {author} {\bibfnamefont {G.~G.}\ \bibnamefont {Adamian}},\ }\href {https://doi.org/10.1103/PhysRevC.107.044601} {\bibfield  {journal} {\bibinfo  {journal} {Phys. Rev. C}\ }\textbf {\bibinfo {volume} {107}},\ \bibinfo {pages} {044601} (\bibinfo {year} {2023})}\BibitemShut {NoStop}%
\bibitem [{\citenamefont {Jolos}\ and\ \citenamefont {Kolganova}(2024)}]{Jolosprc2024}%
  \BibitemOpen
  \bibfield  {author} {\bibinfo {author} {\bibfnamefont {R.~V.}\ \bibnamefont {Jolos}}\ and\ \bibinfo {author} {\bibfnamefont {E.~A.}\ \bibnamefont {Kolganova}},\ }\href {https://doi.org/10.1103/PhysRevC.110.014302} {\bibfield  {journal} {\bibinfo  {journal} {Phys. Rev. C}\ }\textbf {\bibinfo {volume} {110}},\ \bibinfo {pages} {014302} (\bibinfo {year} {2024})}\BibitemShut {NoStop}%
\bibitem [{\citenamefont {S}\ \emph {et~al.}(2024)\citenamefont {S}, \citenamefont {Bozarov}, \citenamefont {Kudiratov},\ and\ \citenamefont {Minkov}}]{MS_2024}%
  \BibitemOpen
  \bibfield  {author} {\bibinfo {author} {\bibfnamefont {N.~M.}\ \bibnamefont {S}}, \bibinfo {author} {\bibfnamefont {O.~A.}\ \bibnamefont {Bozarov}}, \bibinfo {author} {\bibfnamefont {S.~N.}\ \bibnamefont {Kudiratov}},\ and\ \bibinfo {author} {\bibfnamefont {N.}~\bibnamefont {Minkov}},\ }\href {https://doi.org/10.1088/1402-4896/ad6f4d} {\bibfield  {journal} {\bibinfo  {journal} {Phys. Scr.}\ }\textbf {\bibinfo {volume} {99}},\ \bibinfo {pages} {095309} (\bibinfo {year} {2024})}\BibitemShut {NoStop}%
\bibitem [{\citenamefont {Minkov}(2024)}]{Minkov2024}%
  \BibitemOpen
  \bibfield  {author} {\bibinfo {author} {\bibfnamefont {N.}~\bibnamefont {Minkov}},\ }\href {https://doi.org/10.1088/1402-4896/ad4694} {\bibfield  {journal} {\bibinfo  {journal} {Phys. Scr.}\ }\textbf {\bibinfo {volume} {99}},\ \bibinfo {pages} {065303} (\bibinfo {year} {2024})}\BibitemShut {NoStop}%
\bibitem [{\citenamefont {Lu}\ \emph {et~al.}(2012)\citenamefont {Lu}, \citenamefont {Zhao},\ and\ \citenamefont {Zhou}}]{luprc2012r}%
  \BibitemOpen
  \bibfield  {author} {\bibinfo {author} {\bibfnamefont {B.-N.}\ \bibnamefont {Lu}}, \bibinfo {author} {\bibfnamefont {E.-G.}\ \bibnamefont {Zhao}},\ and\ \bibinfo {author} {\bibfnamefont {S.-G.}\ \bibnamefont {Zhou}},\ }\href {https://doi.org/10.1103/PhysRevC.85.011301} {\bibfield  {journal} {\bibinfo  {journal} {Phys. Rev. C}\ }\textbf {\bibinfo {volume} {85}},\ \bibinfo {pages} {011301} (\bibinfo {year} {2012})}\BibitemShut {NoStop}%
\bibitem [{\citenamefont {Zhao}\ \emph {et~al.}(2017)\citenamefont {Zhao}, \citenamefont {Lu}, \citenamefont {Zhao},\ and\ \citenamefont {Zhou}}]{zhaoprc2017}%
  \BibitemOpen
  \bibfield  {author} {\bibinfo {author} {\bibfnamefont {J.}~\bibnamefont {Zhao}}, \bibinfo {author} {\bibfnamefont {B.-N.}\ \bibnamefont {Lu}}, \bibinfo {author} {\bibfnamefont {E.-G.}\ \bibnamefont {Zhao}},\ and\ \bibinfo {author} {\bibfnamefont {S.-G.}\ \bibnamefont {Zhou}},\ }\href {https://doi.org/10.1103/PhysRevC.95.014320} {\bibfield  {journal} {\bibinfo  {journal} {Phys. Rev. C}\ }\textbf {\bibinfo {volume} {95}},\ \bibinfo {pages} {014320} (\bibinfo {year} {2017})}\BibitemShut {NoStop}%
\bibitem [{\citenamefont {Zhao}\ \emph {et~al.}(2015)\citenamefont {Zhao}, \citenamefont {Lu}, \citenamefont {Vretenar}, \citenamefont {Zhao},\ and\ \citenamefont {Zhou}}]{zhaoprc2015}%
  \BibitemOpen
  \bibfield  {author} {\bibinfo {author} {\bibfnamefont {J.}~\bibnamefont {Zhao}}, \bibinfo {author} {\bibfnamefont {B.-N.}\ \bibnamefont {Lu}}, \bibinfo {author} {\bibfnamefont {D.}~\bibnamefont {Vretenar}}, \bibinfo {author} {\bibfnamefont {E.-G.}\ \bibnamefont {Zhao}},\ and\ \bibinfo {author} {\bibfnamefont {S.-G.}\ \bibnamefont {Zhou}},\ }\href {https://doi.org/10.1103/PhysRevC.91.014321} {\bibfield  {journal} {\bibinfo  {journal} {Phys. Rev. C}\ }\textbf {\bibinfo {volume} {91}},\ \bibinfo {pages} {014321} (\bibinfo {year} {2015})}\BibitemShut {NoStop}%
\bibitem [{\citenamefont {Meng}\ \emph {et~al.}(2019)\citenamefont {Meng}, \citenamefont {Lu},\ and\ \citenamefont {Zhou}}]{meng2019-m}%
  \BibitemOpen
  \bibfield  {author} {\bibinfo {author} {\bibfnamefont {X.}~\bibnamefont {Meng}}, \bibinfo {author} {\bibfnamefont {B.}~\bibnamefont {Lu}},\ and\ \bibinfo {author} {\bibfnamefont {S.}~\bibnamefont {Zhou}},\ }\href {https://doi.org/10.1007/s11433-019-9422-1} {\bibfield  {journal} {\bibinfo  {journal} {Sci. China-Phys. Mech. Astron.}\ }\textbf {\bibinfo {volume} {63}},\ \bibinfo {pages} {212011} (\bibinfo {year} {2019})}\BibitemShut {NoStop}%
\bibitem [{\citenamefont {Zhao}\ \emph {et~al.}(2012)\citenamefont {Zhao}, \citenamefont {Lu}, \citenamefont {Zhao},\ and\ \citenamefont {Zhou}}]{zhaoprc2012-m}%
  \BibitemOpen
  \bibfield  {author} {\bibinfo {author} {\bibfnamefont {J.}~\bibnamefont {Zhao}}, \bibinfo {author} {\bibfnamefont {B.-N.}\ \bibnamefont {Lu}}, \bibinfo {author} {\bibfnamefont {E.-G.}\ \bibnamefont {Zhao}},\ and\ \bibinfo {author} {\bibfnamefont {S.-G.}\ \bibnamefont {Zhou}},\ }\href {https://doi.org/10.1103/PhysRevC.86.057304} {\bibfield  {journal} {\bibinfo  {journal} {Phys. Rev. C}\ }\textbf {\bibinfo {volume} {86}},\ \bibinfo {pages} {057304} (\bibinfo {year} {2012})}\BibitemShut {NoStop}%
\bibitem [{\citenamefont {Xu}\ \emph {et~al.}(2022)\citenamefont {Xu}, \citenamefont {Wang}, \citenamefont {Liu}, \citenamefont {Wu}, \citenamefont {Guo}, \citenamefont {Qi}, \citenamefont {Zhao}, \citenamefont {Rohilla}, \citenamefont {Jia}, \citenamefont {Li}, \citenamefont {Zheng}, \citenamefont {Li}, \citenamefont {Han}, \citenamefont {Mu}, \citenamefont {Xiao}, \citenamefont {Wang}, \citenamefont {Sun}, \citenamefont {Li}, \citenamefont {Zhang}, \citenamefont {Wang},\ and\ \citenamefont {Li}}]{XU2022137287}%
  \BibitemOpen
  \bibfield  {author} {\bibinfo {author} {\bibfnamefont {W.}~\bibnamefont {Xu}}, \bibinfo {author} {\bibfnamefont {S.}~\bibnamefont {Wang}}, \bibinfo {author} {\bibfnamefont {C.}~\bibnamefont {Liu}}, \bibinfo {author} {\bibfnamefont {X.}~\bibnamefont {Wu}}, \bibinfo {author} {\bibfnamefont {R.}~\bibnamefont {Guo}}, \bibinfo {author} {\bibfnamefont {B.}~\bibnamefont {Qi}}, \bibinfo {author} {\bibfnamefont {J.}~\bibnamefont {Zhao}}, \bibinfo {author} {\bibfnamefont {A.}~\bibnamefont {Rohilla}}, \bibinfo {author} {\bibfnamefont {H.}~\bibnamefont {Jia}}, \bibinfo {author} {\bibfnamefont {G.}~\bibnamefont {Li}}, \bibinfo {author} {\bibfnamefont {Y.}~\bibnamefont {Zheng}}, \bibinfo {author} {\bibfnamefont {C.}~\bibnamefont {Li}}, \bibinfo {author} {\bibfnamefont {X.}~\bibnamefont {Han}}, \bibinfo {author} {\bibfnamefont {L.}~\bibnamefont {Mu}}, \bibinfo {author} {\bibfnamefont {X.}~\bibnamefont {Xiao}}, \bibinfo {author} {\bibfnamefont {S.}~\bibnamefont {Wang}}, \bibinfo {author} {\bibfnamefont {D.}~\bibnamefont
  {Sun}}, \bibinfo {author} {\bibfnamefont {Z.}~\bibnamefont {Li}}, \bibinfo {author} {\bibfnamefont {Y.}~\bibnamefont {Zhang}}, \bibinfo {author} {\bibfnamefont {C.}~\bibnamefont {Wang}},\ and\ \bibinfo {author} {\bibfnamefont {Y.}~\bibnamefont {Li}},\ }\href {https://doi.org/https://doi.org/10.1016/j.physletb.2022.137287} {\bibfield  {journal} {\bibinfo  {journal} {Phys. Lett. B}\ }\textbf {\bibinfo {volume} {833}},\ \bibinfo {pages} {137287} (\bibinfo {year} {2022})}\BibitemShut {NoStop}%
\bibitem [{\citenamefont {Liu}\ \emph {et~al.}(2016)\citenamefont {Liu}, \citenamefont {Wang}, \citenamefont {Bark}, \citenamefont {Zhang}, \citenamefont {Meng}, \citenamefont {Qi}, \citenamefont {Jones}, \citenamefont {Wyngaardt}, \citenamefont {Zhao}, \citenamefont {Xu}, \citenamefont {Zhou}, \citenamefont {Wang}, \citenamefont {Sun}, \citenamefont {Liu}, \citenamefont {Li}, \citenamefont {Zhang}, \citenamefont {Jia}, \citenamefont {Li}, \citenamefont {Hua}, \citenamefont {Chen}, \citenamefont {Xiao}, \citenamefont {Li}, \citenamefont {Zhu}, \citenamefont {Bucher}, \citenamefont {Dinoko}, \citenamefont {Easton}, \citenamefont {Juh\'asz}, \citenamefont {Kamblawe}, \citenamefont {Khaleel}, \citenamefont {Khumalo}, \citenamefont {Lawrie}, \citenamefont {Lawrie}, \citenamefont {Majola}, \citenamefont {Mullins}, \citenamefont {Murray}, \citenamefont {Ndayishimye}, \citenamefont {Negi}, \citenamefont {Noncolela}, \citenamefont {Ntshangase}, \citenamefont {Nyak\'o}, \citenamefont {Orce}, \citenamefont {Papka},
  \citenamefont {Sharpey-Schafer}, \citenamefont {Shirinda}, \citenamefont {Sithole}, \citenamefont {Stankiewicz},\ and\ \citenamefont {Wiedeking}}]{RN418}%
  \BibitemOpen
  \bibfield  {author} {\bibinfo {author} {\bibfnamefont {C.}~\bibnamefont {Liu}}, \bibinfo {author} {\bibfnamefont {S.~Y.}\ \bibnamefont {Wang}}, \bibinfo {author} {\bibfnamefont {R.~A.}\ \bibnamefont {Bark}}, \bibinfo {author} {\bibfnamefont {S.~Q.}\ \bibnamefont {Zhang}}, \bibinfo {author} {\bibfnamefont {J.}~\bibnamefont {Meng}}, \bibinfo {author} {\bibfnamefont {B.}~\bibnamefont {Qi}}, \bibinfo {author} {\bibfnamefont {P.}~\bibnamefont {Jones}}, \bibinfo {author} {\bibfnamefont {S.~M.}\ \bibnamefont {Wyngaardt}}, \bibinfo {author} {\bibfnamefont {J.}~\bibnamefont {Zhao}}, \bibinfo {author} {\bibfnamefont {C.}~\bibnamefont {Xu}}, \bibinfo {author} {\bibfnamefont {S.-G.}\ \bibnamefont {Zhou}}, \bibinfo {author} {\bibfnamefont {S.}~\bibnamefont {Wang}}, \bibinfo {author} {\bibfnamefont {D.~P.}\ \bibnamefont {Sun}}, \bibinfo {author} {\bibfnamefont {L.}~\bibnamefont {Liu}}, \bibinfo {author} {\bibfnamefont {Z.~Q.}\ \bibnamefont {Li}}, \bibinfo {author} {\bibfnamefont {N.~B.}\ \bibnamefont {Zhang}}, \bibinfo
  {author} {\bibfnamefont {H.}~\bibnamefont {Jia}}, \bibinfo {author} {\bibfnamefont {X.~Q.}\ \bibnamefont {Li}}, \bibinfo {author} {\bibfnamefont {H.}~\bibnamefont {Hua}}, \bibinfo {author} {\bibfnamefont {Q.~B.}\ \bibnamefont {Chen}}, \bibinfo {author} {\bibfnamefont {Z.~G.}\ \bibnamefont {Xiao}}, \bibinfo {author} {\bibfnamefont {H.~J.}\ \bibnamefont {Li}}, \bibinfo {author} {\bibfnamefont {L.~H.}\ \bibnamefont {Zhu}}, \bibinfo {author} {\bibfnamefont {T.~D.}\ \bibnamefont {Bucher}}, \bibinfo {author} {\bibfnamefont {T.}~\bibnamefont {Dinoko}}, \bibinfo {author} {\bibfnamefont {J.}~\bibnamefont {Easton}}, \bibinfo {author} {\bibfnamefont {K.}~\bibnamefont {Juh\'asz}}, \bibinfo {author} {\bibfnamefont {A.}~\bibnamefont {Kamblawe}}, \bibinfo {author} {\bibfnamefont {E.}~\bibnamefont {Khaleel}}, \bibinfo {author} {\bibfnamefont {N.}~\bibnamefont {Khumalo}}, \bibinfo {author} {\bibfnamefont {E.~A.}\ \bibnamefont {Lawrie}}, \bibinfo {author} {\bibfnamefont {J.~J.}\ \bibnamefont {Lawrie}}, \bibinfo {author}
  {\bibfnamefont {S.~N.~T.}\ \bibnamefont {Majola}}, \bibinfo {author} {\bibfnamefont {S.~M.}\ \bibnamefont {Mullins}}, \bibinfo {author} {\bibfnamefont {S.}~\bibnamefont {Murray}}, \bibinfo {author} {\bibfnamefont {J.}~\bibnamefont {Ndayishimye}}, \bibinfo {author} {\bibfnamefont {D.}~\bibnamefont {Negi}}, \bibinfo {author} {\bibfnamefont {S.~P.}\ \bibnamefont {Noncolela}}, \bibinfo {author} {\bibfnamefont {S.~S.}\ \bibnamefont {Ntshangase}}, \bibinfo {author} {\bibfnamefont {B.~M.}\ \bibnamefont {Nyak\'o}}, \bibinfo {author} {\bibfnamefont {J.~N.}\ \bibnamefont {Orce}}, \bibinfo {author} {\bibfnamefont {P.}~\bibnamefont {Papka}}, \bibinfo {author} {\bibfnamefont {J.~F.}\ \bibnamefont {Sharpey-Schafer}}, \bibinfo {author} {\bibfnamefont {O.}~\bibnamefont {Shirinda}}, \bibinfo {author} {\bibfnamefont {P.}~\bibnamefont {Sithole}}, \bibinfo {author} {\bibfnamefont {M.~A.}\ \bibnamefont {Stankiewicz}},\ and\ \bibinfo {author} {\bibfnamefont {M.}~\bibnamefont {Wiedeking}},\ }\href
  {https://doi.org/10.1103/PhysRevLett.116.112501} {\bibfield  {journal} {\bibinfo  {journal} {Phys. Rev. Lett.}\ }\textbf {\bibinfo {volume} {116}},\ \bibinfo {pages} {112501} (\bibinfo {year} {2016})}\BibitemShut {NoStop}%
\bibitem [{\citenamefont {Chen}\ \emph {et~al.}(2016)\citenamefont {Chen}, \citenamefont {Zhao}, \citenamefont {Xu}, \citenamefont {Hua}, \citenamefont {Shneidman}, \citenamefont {Zhou}, \citenamefont {Wu}, \citenamefont {Li}, \citenamefont {Zhang}, \citenamefont {Li}, \citenamefont {Liang}, \citenamefont {Meng}, \citenamefont {Xu}, \citenamefont {Qi}, \citenamefont {Ye}, \citenamefont {Jiang}, \citenamefont {Cheng}, \citenamefont {He}, \citenamefont {Sun}, \citenamefont {Han}, \citenamefont {Niu}, \citenamefont {Li}, \citenamefont {Li}, \citenamefont {Wang}, \citenamefont {Wu}, \citenamefont {Li}, \citenamefont {Zhou}, \citenamefont {Hu}, \citenamefont {Zhang}, \citenamefont {Li}, \citenamefont {He}, \citenamefont {Zheng}, \citenamefont {Li}, \citenamefont {Li}, \citenamefont {Wu}, \citenamefont {Luo},\ and\ \citenamefont {Zhong}}]{Chen021301}%
  \BibitemOpen
  \bibfield  {author} {\bibinfo {author} {\bibfnamefont {X.~C.}\ \bibnamefont {Chen}}, \bibinfo {author} {\bibfnamefont {J.}~\bibnamefont {Zhao}}, \bibinfo {author} {\bibfnamefont {C.}~\bibnamefont {Xu}}, \bibinfo {author} {\bibfnamefont {H.}~\bibnamefont {Hua}}, \bibinfo {author} {\bibfnamefont {T.~M.}\ \bibnamefont {Shneidman}}, \bibinfo {author} {\bibfnamefont {S.~G.}\ \bibnamefont {Zhou}}, \bibinfo {author} {\bibfnamefont {X.~G.}\ \bibnamefont {Wu}}, \bibinfo {author} {\bibfnamefont {X.~Q.}\ \bibnamefont {Li}}, \bibinfo {author} {\bibfnamefont {S.~Q.}\ \bibnamefont {Zhang}}, \bibinfo {author} {\bibfnamefont {Z.~H.}\ \bibnamefont {Li}}, \bibinfo {author} {\bibfnamefont {W.~Y.}\ \bibnamefont {Liang}}, \bibinfo {author} {\bibfnamefont {J.}~\bibnamefont {Meng}}, \bibinfo {author} {\bibfnamefont {F.~R.}\ \bibnamefont {Xu}}, \bibinfo {author} {\bibfnamefont {B.}~\bibnamefont {Qi}}, \bibinfo {author} {\bibfnamefont {Y.~L.}\ \bibnamefont {Ye}}, \bibinfo {author} {\bibfnamefont {D.~X.}\ \bibnamefont {Jiang}},
  \bibinfo {author} {\bibfnamefont {Y.~Y.}\ \bibnamefont {Cheng}}, \bibinfo {author} {\bibfnamefont {C.}~\bibnamefont {He}}, \bibinfo {author} {\bibfnamefont {J.~J.}\ \bibnamefont {Sun}}, \bibinfo {author} {\bibfnamefont {R.}~\bibnamefont {Han}}, \bibinfo {author} {\bibfnamefont {C.~Y.}\ \bibnamefont {Niu}}, \bibinfo {author} {\bibfnamefont {C.~G.}\ \bibnamefont {Li}}, \bibinfo {author} {\bibfnamefont {P.~J.}\ \bibnamefont {Li}}, \bibinfo {author} {\bibfnamefont {C.~G.}\ \bibnamefont {Wang}}, \bibinfo {author} {\bibfnamefont {H.~Y.}\ \bibnamefont {Wu}}, \bibinfo {author} {\bibfnamefont {Z.~H.}\ \bibnamefont {Li}}, \bibinfo {author} {\bibfnamefont {H.}~\bibnamefont {Zhou}}, \bibinfo {author} {\bibfnamefont {S.~P.}\ \bibnamefont {Hu}}, \bibinfo {author} {\bibfnamefont {H.~Q.}\ \bibnamefont {Zhang}}, \bibinfo {author} {\bibfnamefont {G.~S.}\ \bibnamefont {Li}}, \bibinfo {author} {\bibfnamefont {C.~Y.}\ \bibnamefont {He}}, \bibinfo {author} {\bibfnamefont {Y.}~\bibnamefont {Zheng}}, \bibinfo {author}
  {\bibfnamefont {C.~B.}\ \bibnamefont {Li}}, \bibinfo {author} {\bibfnamefont {H.~W.}\ \bibnamefont {Li}}, \bibinfo {author} {\bibfnamefont {Y.~H.}\ \bibnamefont {Wu}}, \bibinfo {author} {\bibfnamefont {P.~W.}\ \bibnamefont {Luo}},\ and\ \bibinfo {author} {\bibfnamefont {J.}~\bibnamefont {Zhong}},\ }\href {https://doi.org/10.1103/PhysRevC.94.021301} {\bibfield  {journal} {\bibinfo  {journal} {Phys. Rev. C}\ }\textbf {\bibinfo {volume} {94}},\ \bibinfo {pages} {021301} (\bibinfo {year} {2016})}\BibitemShut {NoStop}%
\bibitem [{\citenamefont {Wang}\ \emph {et~al.}(2022{\natexlab{d}})\citenamefont {Wang}, \citenamefont {Chen},\ and\ \citenamefont {Zhang}}]{Wang044316}%
  \BibitemOpen
  \bibfield  {author} {\bibinfo {author} {\bibfnamefont {Y.~Y.}\ \bibnamefont {Wang}}, \bibinfo {author} {\bibfnamefont {Q.~B.}\ \bibnamefont {Chen}},\ and\ \bibinfo {author} {\bibfnamefont {S.~Q.}\ \bibnamefont {Zhang}},\ }\href {https://doi.org/10.1103/PhysRevC.105.044316} {\bibfield  {journal} {\bibinfo  {journal} {Phys. Rev. C}\ }\textbf {\bibinfo {volume} {105}},\ \bibinfo {pages} {044316} (\bibinfo {year} {2022}{\natexlab{d}})}\BibitemShut {NoStop}%
\bibitem [{\citenamefont {Lu}\ \emph {et~al.}(2011)\citenamefont {Lu}, \citenamefont {Zhao},\ and\ \citenamefont {Zhou}}]{Lu014328}%
  \BibitemOpen
  \bibfield  {author} {\bibinfo {author} {\bibfnamefont {B.-N.}\ \bibnamefont {Lu}}, \bibinfo {author} {\bibfnamefont {E.-G.}\ \bibnamefont {Zhao}},\ and\ \bibinfo {author} {\bibfnamefont {S.-G.}\ \bibnamefont {Zhou}},\ }\href {https://doi.org/10.1103/PhysRevC.84.014328} {\bibfield  {journal} {\bibinfo  {journal} {Phys. Rev. C}\ }\textbf {\bibinfo {volume} {84}},\ \bibinfo {pages} {014328} (\bibinfo {year} {2011})}\BibitemShut {NoStop}%
\bibitem [{\citenamefont {Lu}\ \emph {et~al.}(2014{\natexlab{b}})\citenamefont {Lu}, \citenamefont {Hiyama}, \citenamefont {Sagawa},\ and\ \citenamefont {Zhou}}]{Lu044307}%
  \BibitemOpen
  \bibfield  {author} {\bibinfo {author} {\bibfnamefont {B.-N.}\ \bibnamefont {Lu}}, \bibinfo {author} {\bibfnamefont {E.}~\bibnamefont {Hiyama}}, \bibinfo {author} {\bibfnamefont {H.}~\bibnamefont {Sagawa}},\ and\ \bibinfo {author} {\bibfnamefont {S.-G.}\ \bibnamefont {Zhou}},\ }\href {https://doi.org/10.1103/PhysRevC.89.044307} {\bibfield  {journal} {\bibinfo  {journal} {Phys. Rev. C}\ }\textbf {\bibinfo {volume} {89}},\ \bibinfo {pages} {044307} (\bibinfo {year} {2014}{\natexlab{b}})}\BibitemShut {NoStop}%
\bibitem [{\citenamefont {Rong}\ \emph {et~al.}(2020)\citenamefont {Rong}, \citenamefont {Zhao},\ and\ \citenamefont {Zhou}}]{RONG2020135533}%
  \BibitemOpen
  \bibfield  {author} {\bibinfo {author} {\bibfnamefont {Y.-T.}\ \bibnamefont {Rong}}, \bibinfo {author} {\bibfnamefont {P.}~\bibnamefont {Zhao}},\ and\ \bibinfo {author} {\bibfnamefont {S.-G.}\ \bibnamefont {Zhou}},\ }\href {https://doi.org/https://doi.org/10.1016/j.physletb.2020.135533} {\bibfield  {journal} {\bibinfo  {journal} {Physics Letters B}\ }\textbf {\bibinfo {volume} {807}},\ \bibinfo {pages} {135533} (\bibinfo {year} {2020})}\BibitemShut {NoStop}%
\bibitem [{\citenamefont {Rong}\ \emph {et~al.}(2021)\citenamefont {Rong}, \citenamefont {Tu},\ and\ \citenamefont {Zhou}}]{Rong054321}%
  \BibitemOpen
  \bibfield  {author} {\bibinfo {author} {\bibfnamefont {Y.-T.}\ \bibnamefont {Rong}}, \bibinfo {author} {\bibfnamefont {Z.-H.}\ \bibnamefont {Tu}},\ and\ \bibinfo {author} {\bibfnamefont {S.-G.}\ \bibnamefont {Zhou}},\ }\href {https://doi.org/10.1103/PhysRevC.104.054321} {\bibfield  {journal} {\bibinfo  {journal} {Phys. Rev. C}\ }\textbf {\bibinfo {volume} {104}},\ \bibinfo {pages} {054321} (\bibinfo {year} {2021})}\BibitemShut {NoStop}%
\bibitem [{\citenamefont {Chen}\ \emph {et~al.}(2021)\citenamefont {Chen}, \citenamefont {Sun}, \citenamefont {Li},\ and\ \citenamefont {Sun}}]{RN956}%
  \BibitemOpen
  \bibfield  {author} {\bibinfo {author} {\bibfnamefont {C.}~\bibnamefont {Chen}}, \bibinfo {author} {\bibfnamefont {Q.-K.}\ \bibnamefont {Sun}}, \bibinfo {author} {\bibfnamefont {Y.-X.}\ \bibnamefont {Li}},\ and\ \bibinfo {author} {\bibfnamefont {T.-T.}\ \bibnamefont {Sun}},\ }\href {https://doi.org/10.1007/s11433-021-1721-1} {\bibfield  {journal} {\bibinfo  {journal} {Sci. China-Phys. Mech. Astron.}\ }\textbf {\bibinfo {volume} {64}},\ \bibinfo {pages} {282011} (\bibinfo {year} {2021})}\BibitemShut {NoStop}%
\bibitem [{\citenamefont {Sun}\ \emph {et~al.}(2022)\citenamefont {Sun}, \citenamefont {Sun}, \citenamefont {Zhang}, \citenamefont {Zhang},\ and\ \citenamefont {Chen}}]{Sun6153}%
  \BibitemOpen
  \bibfield  {author} {\bibinfo {author} {\bibfnamefont {Q.-K.}\ \bibnamefont {Sun}}, \bibinfo {author} {\bibfnamefont {T.-T.}\ \bibnamefont {Sun}}, \bibinfo {author} {\bibfnamefont {W.}~\bibnamefont {Zhang}}, \bibinfo {author} {\bibfnamefont {S.-S.}\ \bibnamefont {Zhang}},\ and\ \bibinfo {author} {\bibfnamefont {C.}~\bibnamefont {Chen}},\ }\href {https://doi.org/10.1088/1674-1137/ac6153} {\bibfield  {journal} {\bibinfo  {journal} {Chin. Phys. C}\ }\textbf {\bibinfo {volume} {46}},\ \bibinfo {pages} {074106} (\bibinfo {year} {2022})}\BibitemShut {NoStop}%
\bibitem [{\citenamefont {Tian}\ \emph {et~al.}(2009{\natexlab{a}})\citenamefont {Tian}, \citenamefont {Ma},\ and\ \citenamefont {Ring}}]{TIAN200944}%
  \BibitemOpen
  \bibfield  {author} {\bibinfo {author} {\bibfnamefont {Y.}~\bibnamefont {Tian}}, \bibinfo {author} {\bibfnamefont {Z.}~\bibnamefont {Ma}},\ and\ \bibinfo {author} {\bibfnamefont {P.}~\bibnamefont {Ring}},\ }\href {https://doi.org/https://doi.org/10.1016/j.physletb.2009.04.067} {\bibfield  {journal} {\bibinfo  {journal} {Phys. Lett. B}\ }\textbf {\bibinfo {volume} {676}},\ \bibinfo {pages} {44} (\bibinfo {year} {2009}{\natexlab{a}})}\BibitemShut {NoStop}%
\bibitem [{\citenamefont {Tian}\ \emph {et~al.}(2009{\natexlab{b}})\citenamefont {Tian}, \citenamefont {Ma},\ and\ \citenamefont {Ring}}]{PRC09tian}%
  \BibitemOpen
  \bibfield  {author} {\bibinfo {author} {\bibfnamefont {Y.}~\bibnamefont {Tian}}, \bibinfo {author} {\bibfnamefont {Z.-y.}\ \bibnamefont {Ma}},\ and\ \bibinfo {author} {\bibfnamefont {P.}~\bibnamefont {Ring}},\ }\href {https://doi.org/10.1103/PhysRevC.79.064301} {\bibfield  {journal} {\bibinfo  {journal} {Phys. Rev. C}\ }\textbf {\bibinfo {volume} {79}},\ \bibinfo {pages} {064301} (\bibinfo {year} {2009}{\natexlab{b}})}\BibitemShut {NoStop}%
\bibitem [{\citenamefont {Tian}\ \emph {et~al.}(2009{\natexlab{c}})\citenamefont {Tian}, \citenamefont {Ma},\ and\ \citenamefont {Ring}}]{PRC09tian-1}%
  \BibitemOpen
  \bibfield  {author} {\bibinfo {author} {\bibfnamefont {Y.}~\bibnamefont {Tian}}, \bibinfo {author} {\bibfnamefont {Z.-y.}\ \bibnamefont {Ma}},\ and\ \bibinfo {author} {\bibfnamefont {P.}~\bibnamefont {Ring}},\ }\href {https://doi.org/10.1103/PhysRevC.80.024313} {\bibfield  {journal} {\bibinfo  {journal} {Phys. Rev. C}\ }\textbf {\bibinfo {volume} {80}},\ \bibinfo {pages} {024313} (\bibinfo {year} {2009}{\natexlab{c}})}\BibitemShut {NoStop}%
\bibitem [{\citenamefont {Zhao}\ \emph {et~al.}(2010)\citenamefont {Zhao}, \citenamefont {Li}, \citenamefont {Yao},\ and\ \citenamefont {Meng}}]{Zhaopw10}%
  \BibitemOpen
  \bibfield  {author} {\bibinfo {author} {\bibfnamefont {P.~W.}\ \bibnamefont {Zhao}}, \bibinfo {author} {\bibfnamefont {Z.~P.}\ \bibnamefont {Li}}, \bibinfo {author} {\bibfnamefont {J.~M.}\ \bibnamefont {Yao}},\ and\ \bibinfo {author} {\bibfnamefont {J.}~\bibnamefont {Meng}},\ }\href {https://doi.org/10.1103/PhysRevC.82.054319} {\bibfield  {journal} {\bibinfo  {journal} {Phys. Rev. C}\ }\textbf {\bibinfo {volume} {82}},\ \bibinfo {pages} {054319} (\bibinfo {year} {2010})}\BibitemShut {NoStop}%
\bibitem [{\citenamefont {Yao}\ \emph {et~al.}(2009)\citenamefont {Yao}, \citenamefont {Meng}, \citenamefont {Ring},\ and\ \citenamefont {Arteaga}}]{yao09}%
  \BibitemOpen
  \bibfield  {author} {\bibinfo {author} {\bibfnamefont {J.~M.}\ \bibnamefont {Yao}}, \bibinfo {author} {\bibfnamefont {J.}~\bibnamefont {Meng}}, \bibinfo {author} {\bibfnamefont {P.}~\bibnamefont {Ring}},\ and\ \bibinfo {author} {\bibfnamefont {D.~P.}\ \bibnamefont {Arteaga}},\ }\href {https://doi.org/10.1103/PhysRevC.79.044312} {\bibfield  {journal} {\bibinfo  {journal} {Phys. Rev. C}\ }\textbf {\bibinfo {volume} {79}},\ \bibinfo {pages} {044312} (\bibinfo {year} {2009})}\BibitemShut {NoStop}%
\bibitem [{\citenamefont {Yao}\ \emph {et~al.}(2010)\citenamefont {Yao}, \citenamefont {Meng}, \citenamefont {Ring},\ and\ \citenamefont {Vretenar}}]{yao044311}%
  \BibitemOpen
  \bibfield  {author} {\bibinfo {author} {\bibfnamefont {J.~M.}\ \bibnamefont {Yao}}, \bibinfo {author} {\bibfnamefont {J.}~\bibnamefont {Meng}}, \bibinfo {author} {\bibfnamefont {P.}~\bibnamefont {Ring}},\ and\ \bibinfo {author} {\bibfnamefont {D.}~\bibnamefont {Vretenar}},\ }\href {https://doi.org/10.1103/PhysRevC.81.044311} {\bibfield  {journal} {\bibinfo  {journal} {Phys. Rev. C}\ }\textbf {\bibinfo {volume} {81}},\ \bibinfo {pages} {044311} (\bibinfo {year} {2010})}\BibitemShut {NoStop}%
\bibitem [{\citenamefont {Rudolph}\ \emph {et~al.}(1998)\citenamefont {Rudolph}, \citenamefont {Baktash}, \citenamefont {Dobaczewski}, \citenamefont {Nazarewicz}, \citenamefont {Satula}, \citenamefont {Brinkman}, \citenamefont {Devlin}, \citenamefont {Jin}, \citenamefont {LaFosse}, \citenamefont {Riedinger}, \citenamefont {Sarantites},\ and\ \citenamefont {Yu}}]{RN618}%
  \BibitemOpen
  \bibfield  {author} {\bibinfo {author} {\bibfnamefont {D.}~\bibnamefont {Rudolph}}, \bibinfo {author} {\bibfnamefont {C.}~\bibnamefont {Baktash}}, \bibinfo {author} {\bibfnamefont {J.}~\bibnamefont {Dobaczewski}}, \bibinfo {author} {\bibfnamefont {W.}~\bibnamefont {Nazarewicz}}, \bibinfo {author} {\bibfnamefont {W.}~\bibnamefont {Satula}}, \bibinfo {author} {\bibfnamefont {M.~J.}\ \bibnamefont {Brinkman}}, \bibinfo {author} {\bibfnamefont {M.}~\bibnamefont {Devlin}}, \bibinfo {author} {\bibfnamefont {H.~Q.}\ \bibnamefont {Jin}}, \bibinfo {author} {\bibfnamefont {D.~R.}\ \bibnamefont {LaFosse}}, \bibinfo {author} {\bibfnamefont {L.~L.}\ \bibnamefont {Riedinger}}, \bibinfo {author} {\bibfnamefont {D.~G.}\ \bibnamefont {Sarantites}},\ and\ \bibinfo {author} {\bibfnamefont {C.~H.}\ \bibnamefont {Yu}},\ }\href {https://doi.org/DOI 10.1103/PhysRevLett.80.3018} {\bibfield  {journal} {\bibinfo  {journal} {Phys. Rev. Lett.}\ }\textbf {\bibinfo {volume} {80}},\ \bibinfo {pages} {3018} (\bibinfo {year}
  {1998})}\BibitemShut {NoStop}%
\bibitem [{\citenamefont {Nik\ifmmode \check{s}\else \v{s}\fi{}i\ifmmode~\acute{c}\else \'{c}\fi{}}\ \emph {et~al.}(2006)\citenamefont {Nik\ifmmode \check{s}\else \v{s}\fi{}i\ifmmode~\acute{c}\else \'{c}\fi{}}, \citenamefont {Vretenar},\ and\ \citenamefont {Ring}}]{RN769}%
  \BibitemOpen
  \bibfield  {author} {\bibinfo {author} {\bibfnamefont {T.}~\bibnamefont {Nik\ifmmode \check{s}\else \v{s}\fi{}i\ifmmode~\acute{c}\else \'{c}\fi{}}}, \bibinfo {author} {\bibfnamefont {D.}~\bibnamefont {Vretenar}},\ and\ \bibinfo {author} {\bibfnamefont {P.}~\bibnamefont {Ring}},\ }\href {https://doi.org/10.1103/PhysRevC.73.034308} {\bibfield  {journal} {\bibinfo  {journal} {Phys. Rev. C}\ }\textbf {\bibinfo {volume} {73}},\ \bibinfo {pages} {034308} (\bibinfo {year} {2006})}\BibitemShut {NoStop}%
\bibitem [{\citenamefont {Nann}\ and\ \citenamefont {Benenson}(1974)}]{ni561974}%
  \BibitemOpen
  \bibfield  {author} {\bibinfo {author} {\bibfnamefont {H.}~\bibnamefont {Nann}}\ and\ \bibinfo {author} {\bibfnamefont {W.}~\bibnamefont {Benenson}},\ }\href {https://doi.org/10.1103/PhysRevC.10.1880} {\bibfield  {journal} {\bibinfo  {journal} {Phys. Rev. C}\ }\textbf {\bibinfo {volume} {10}},\ \bibinfo {pages} {1880} (\bibinfo {year} {1974})}\BibitemShut {NoStop}%
\bibitem [{\citenamefont {Zhao}\ \emph {et~al.}(2016)\citenamefont {Zhao}, \citenamefont {Ring},\ and\ \citenamefont {Meng}}]{zhao041301}%
  \BibitemOpen
  \bibfield  {author} {\bibinfo {author} {\bibfnamefont {P.~W.}\ \bibnamefont {Zhao}}, \bibinfo {author} {\bibfnamefont {P.}~\bibnamefont {Ring}},\ and\ \bibinfo {author} {\bibfnamefont {J.}~\bibnamefont {Meng}},\ }\href {https://doi.org/10.1103/PhysRevC.94.041301} {\bibfield  {journal} {\bibinfo  {journal} {Phys. Rev. C}\ }\textbf {\bibinfo {volume} {94}},\ \bibinfo {pages} {041301} (\bibinfo {year} {2016})}\BibitemShut {NoStop}%
\bibitem [{\citenamefont {Sun}\ and\ \citenamefont {Zhou}(2021)}]{sun064319}%
  \BibitemOpen
  \bibfield  {author} {\bibinfo {author} {\bibfnamefont {X.-X.}\ \bibnamefont {Sun}}\ and\ \bibinfo {author} {\bibfnamefont {S.-G.}\ \bibnamefont {Zhou}},\ }\href {https://doi.org/10.1103/PhysRevC.104.064319} {\bibfield  {journal} {\bibinfo  {journal} {Phys. Rev. C}\ }\textbf {\bibinfo {volume} {104}},\ \bibinfo {pages} {064319} (\bibinfo {year} {2021})}\BibitemShut {NoStop}%
\bibitem [{\citenamefont {Zhou}\ \emph {et~al.}(2024{\natexlab{b}})\citenamefont {Zhou}, \citenamefont {Wu},\ and\ \citenamefont {Yao}}]{zhou034305}%
  \BibitemOpen
  \bibfield  {author} {\bibinfo {author} {\bibfnamefont {E.~F.}\ \bibnamefont {Zhou}}, \bibinfo {author} {\bibfnamefont {X.~Y.}\ \bibnamefont {Wu}},\ and\ \bibinfo {author} {\bibfnamefont {J.~M.}\ \bibnamefont {Yao}},\ }\href {https://doi.org/10.1103/PhysRevC.109.034305} {\bibfield  {journal} {\bibinfo  {journal} {Phys. Rev. C}\ }\textbf {\bibinfo {volume} {109}},\ \bibinfo {pages} {034305} (\bibinfo {year} {2024}{\natexlab{b}})}\BibitemShut {NoStop}%
\end{thebibliography}%
\end{document}